\documentclass[preprint,aps,prb,nofootinbib,groupedaddress,endfloats*,floatfix,superscriptaddress,pdflatex]{revtex4}
\pdfoutput=1 

\usepackage{graphics}
\usepackage{graphicx}
\usepackage{amssymb}
\usepackage{amsmath}
\usepackage{amsfonts}
\usepackage{color}
\usepackage{setspace}
\usepackage{multirow}
\usepackage{lscape}
\usepackage{slashbox}
\usepackage{array}

\begin{document}

\title{Structure map of AB$_2$ type 2D materials by high-throughput DFT calculations}

\author{Masahiro Fukuda}
\affiliation{Institute for Solid State Physics, The University of Tokyo,
5-1-5 Kashiwanoha, Kashiwa, Chiba 277-8581, Japan}
\author{Jingning Zhang}
\affiliation{Department of Physics, University of Science
and Technology of China, Hefei, Anhui 230026, China}
\author{Yung-Ting Lee}
\affiliation{Institute for Solid State Physics, The University of Tokyo,
5-1-5 Kashiwanoha, Kashiwa, Chiba 277-8581, Japan}
\author{Taisuke Ozaki}
\affiliation{Institute for Solid State Physics, The University of Tokyo,
5-1-5 Kashiwanoha, Kashiwa, Chiba 277-8581, Japan}
\email{masahiro.fukuda@issp.u-tokyo.ac.jp}

\date{\today}

\begin{abstract}
By high-throughput calculations based on the density functional theory,
we construct a structure map for AB$_2$ type monolayers of 3844 compounds
which are all the combinations of 62 elements selected from the periodic table.
The structure map and its web version
(www.openmx-square.org/2d-ab2/)
provide comprehensive structural trends of the 3844 compounds
in two dimensional (2D) structures,
and predict correctly structures of most of existing 2D compounds
such as transition metal dichalcogenides and MXenes having 1T or 1H type structures.
We also summarize all the families of 1T/1H type AB$_2$ monolayers for each combination of groups
in the periodic table on the basis of our structure map,
and propose new types of structures such as a memory structure,
which may be a candidate material for data storage applications with an extremely high areal density.
In addition, planar and distorted planar structures
and other geometrically characteristic structures are found through the high-throughput calculations.
These characteristic structures might give new viewpoints and directions to search unknown 2D materials.
Our structure map and database will promote efforts towards synthesizing undiscovered 2D materials experimentally
and investigating properties of the new structures theoretically.
\end{abstract}


\maketitle
\section{Introduction}

Development of both experiments and theoretical calculations 
has brought recent rapid research progress of two-dimensional (2D) materials~\cite{Miro_2014,Zhou_2019,Matsuda_2018},
which have a vast diversity of chemical and physical properties.
Through these recent researches, 2D materials have been shown to be metals, semimetals, semiconductors, insulators, superconductors, and exhibit high magnetoresistance~\cite{YFeng_2018} and a wide range of electronic and thermal conductivities~\cite{Liao_2015,Yoshida_2016}. 
Owing to such a rich variety of electronic, thermal, and magnetic properties,
2D materials are expected to be good candidates for
a wide range of future applications in electronic, opto-electronic~\cite{Wang_2012}, photonic~\cite{Mak_2016}, spintronic~\cite{Zibouche_2014,Li_2014}, and valleytronic devices~\cite{Ezawa_2013,Zeng_2012},
batteries~\cite{Chang_2011,Yoo_2008}, solar cells~\cite{Das_2019}, sensors~\cite{Schedin_2007}, and energy storage~\cite{Ataca_2008,Zhou_2010}.

It is worth mentioning that such discoveries of new 2D materials have been supported by
improvement of measurement technology and computational methods for theoretical calculations.
For example,
Angle-Resolved PhotoEmission Spectroscopy (ARPES)~\cite{Damascelli_2003,CCLee_2013,CCLee_2018},
X-ray Photoelectron Spectroscopy (XPS)~\cite{Fadley_1968,Ozaki_2017},
Scanning Tunneling Microscope (STM)~\cite{Binnig_1982,Tersoff_1985},
and Atomic Force Microscope (AFM)~\cite{Feng_2018},
have made a paradigm shift from researches about bulk feature in solids
to investigation of surface science,
where the electronic structure calculations
have been becoming indispensable tools
to elucidate structures and chemical and physical properties on the surfaces of solids.

Many 2D materials have already been investigated theoretically and experimentally.
As specific examples of mono-elemental 2D materials,
semimetals (graphene~\cite{Geim_2007}, silicene~\cite{Takeda_1994,Augray_2010,Lalmi_2010,Fleurence_2012,Lee_2014,Lee_2017,Feng_2018}, germanene~\cite{Bianco_2013,Davila_2014}, and antimonene\cite{Shao_2018}),
metals ($\beta_{12}$-borophene sheet~\cite{Mannix_2015,Feng_2015} and stanene~\cite{Zhu_2015}),
semiconductors (phosphorene~\cite{Liu_2015}),
and topological insulator (bismuthene~\cite{Reis_2017})
have already been synthesized experimentally.
In addition, h-BN~\cite{Lin_2010} is known as an insulator of AB type monolayer,
and AgSi~\cite{Takahashi_1988,Aizawa_1999}, Cu$_2$Si~\cite{Feng2_2015,Yang_2015}, and In$_2$Si~\cite{Hashimoto_2009} are known as metal layer structures
synthesized on a Si(111) surface.

Furthermore, various kinds of monolayers of transition metal (A=Mo, W, etc.) 
dichalcogenides (B=S, Se or Te) with a formula AB$_2$ (TMDCs)
have attracted attentions owing to the development of
the liquid exfoliation methods and the chemical vapour deposition methods~\cite{Zhou_2018}.
The diversity of the physical properties originating from their compositions
also contributes to attract researchers in applied physics.
The transition metal dioxides (TMDOs) nanosheet including CoO$_2$~\cite{Takada_2003}, MnO$_2$~\cite{Omomo_2003}, and RuO$_2$~\cite{Shikano_2004},
and MXenes (transition metal carbides/nitrides)~\cite{Naguib_2014,Tanga_2018,Zhu_2017} have already been reported as new families
of 2D materials produced by the intercalation and the surface-terminating technology.
Unlike honeycomb planar-like mono-elemental 2D materials,
most of the structures of the TMDCs, TMDOs, and MXenes are classified into the 1T/1H structures;
the T and H stand for trigonal and hexagonal, respectively,
and the numbers indicate the number of layers in the unit cell.
Moreover, the unique porous structures of the carbon nitride monolayers including CN, C$_2$N, and C$_3$N$_4$
have been reported as candidates for porous membranes in water treatment technologies~\cite{Zhou_2019}.

As mentioned above,
many 2D materials have already been synthesized or exfoliated experimentally.
However,
since the combination of the elements is diverse even only for the AB$_2$ composition,
it can be considered that there is a sufficient room for exploring unknown stable monolayer structures
by computational simulation.
As for three-dimensional crystals,
in 1980s Zunger, Villars, and Pettifor reported 
structure maps which were constructed to predict 3D crystal structures
for specific element combinations such as AB and AB$_2$~\cite{Zunger_1980,Villars_1983,Pettifor_1984,Pettifor_1988}.
These structure maps contributed to comprehensively understand structural trends
and give an initial guide in the search for new 3D materials.
Unlike bulk materials,
it might be possible to find a variety of unknown structures,
since 2D materials can exist even in a thermodynamically metastable state.
Therefore, the constructions of the structure maps for 2D materials 
with specific element combinations such as AB and AB$_2$
are also desired to contribute to new synthesis experiments and
theoretical explorations of unique physical properties.
Although some studies about high-throughput calculations and databases of
2D materials have already been reported due to the improvement of
computer performance and the development of electronic structure calculation program codes,
their studies are still few.
For example,
databases of AB$_2$ TMDCs and TMDOs
by using DFT calculation have been reported~\cite{Ataca_2012,XZhang_2018,Rasmussen_2015}.
The most recent study about database construction for exploration
of 2D materials by using high-throughput calculations
has been reported by S.~Haastrup \textit{et al.}~\cite{Haastrup_2018}
and M.~Ashton \textit{et al.}~\cite{Ashton_2017}.
Furthermore, development of synthesis technologies such as
a molten-salt-assisted chemical vapour deposition method
and observation technologies such as the scanning transmission electron microscope
made it possible to construct a library of the TMDCs experimentally~\cite{Zhou_2018}.
Such databases can be a guide to synthesize new materials experimentally
and to explore new chemical and physical properties theoretically.
In addition, it is expected that
such huge databases can also be utilized for data mining and machine learning techniques
to explore suitable structures and properties for new applications.
Thus,
structural exploration of AB$_2$ type 2D materials
can be regarded as a good example to demonstrate high-throughput calculations,
since it is possible to compare computational results directly with experimental results.

In this study,
we focus on constructing a structure map for AB$_2$ type 2D materials
on the basis of the high-throughput DFT calculations
to discuss geometrical atomic structures themselves.
Our main purpose of this study is to find AB$_2$ type structures
which are possible to be 1T or 1H such as TMDCs
and also to find new types and families of 2D structures.
This paper is organized as follows.
In Sec.~\ref{sec:Computational_details},
the methods and workflow for the high-throughput DFT calculations are outlined.
The structure map and the detailed classification by using the space-group
are discussed in Secs.~\ref{sec:map} and \ref{sec:spg}.
The results of 1T/1H structures such as TMDCs (Sec.~\ref{sec:TMDC_TMDO}), TMDOs (Sec.~\ref{sec:TMDC_TMDO}),
metal dihalides (Sec.~\ref{sec:dihalides}),
MXenes and BiXenes (Sec.~\ref{sec:MXene}), and other families of 1T/1H structures (Sec.~\ref{sec:other_1T_1H})
are discussed comparing to the reported theoretical calculations and experimental results in Sec.~\ref{sec:1T_1H}.
All the families of 1T/1H structures for each combination of groups in the periodic table
on the basis of our structure map are summarized in Sec.~\ref{sec:other_1T_1H}.
Planar and distorted planar structures and memory structures,
which can be candidates for data storage applications,
are also discussed in Secs.~\ref{sec:Planar} and \ref{sec:Memory}.
The other characteristic structures we found in this study are discussed in Sec.~\ref{sec:characteristic}.
Finally, Sec.~\ref{sec:Conclusions} is devoted to our conclusions.



\section{Computational details}
\label{sec:Computational_details}

\begin{figure}[htb]
\centering
\includegraphics[width=0.95\linewidth]{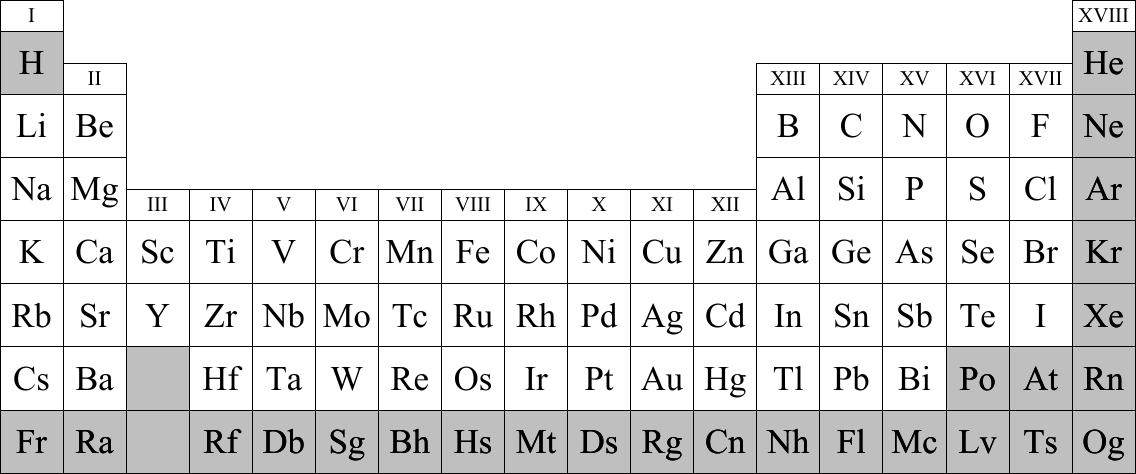}
\caption{Periodic table. The uncolored 62 elements were used for our high-throughput calculations.}\label{fig:periodic_table}
\end{figure}

High-throughput calculations based on DFT were performed to construct a structure map for AB$_2$ type monolayers.
The combinations of the atoms A and B were chosen from elements in the periodic table,
excluding hydrogen atom, noble gases, lanthanoids and actinoids (see Fig.~\ref{fig:periodic_table}).
Therefore, geometry relaxations and variable cell optimizations were carried out for 3844 compounds (= 62 elements $\times$ 62 elements)
including mono elemental substances (A=B cases).
Here, we note that, in our paper, the names of compounds are sometimes expressed simply as AB$_2$ in order
instead of conventional expressions such as Ti$_2$C (dititanium carbide).
The DFT calculations within a generalized gradient approximation (GGA) \cite{Kohn_2018,Perdew_1996} 
were performed using the
OpenMX code \cite{openmx} based on norm-conserving pseudopotentials generated
with multireference energies \cite{Morrison_1993} and optimized pseudoatomic
basis functions\cite{Ozaki_2003}.
The basis sets we used are listed in Appendix~\ref{sec:List_basis}.
For example, Fe6.0H-s3p2d1 means that three, two, and one optimized radial functions
were allocated for the $s$, $p$, and $d$ orbitals, respectively for Fe atoms with the ``hard'' pseudopotential,
and the cutoff radius of 6 Bohr was chosen.
The qualities of basis functions and pseudopotentials were carefully
benchmarked by the delta gauge method~\cite{Lejaeghere_2016}
to ensure accuracy of our calculations.
An electronic temperature of 700 K is used to count the number
of electrons by the Fermi-Dirac function.
The regular mesh of 240 Ry in real space was used for the numerical
integrations and for the solution of the Poisson equation~\cite{Ozaki_2005}.
A $5 \times 5 \times 1$ mesh of k points was adopted.
Cell vectors and internal coordinates are simultaneously optimized
without any constraint by using a combination scheme of
the rational function (RF) method~\cite{Banerjee_1985} 
and the direct inversion iterative sub-space (DIIS) method~\cite{Csaszar_1984}
with a BFGS update~\cite{Broyden_1970} for the approximate Hessian. 
The force on each atom was relaxed to be less than 0.0005 Hartree/Bohr.

\begin{figure}[htb]
\includegraphics[width=0.95\linewidth]{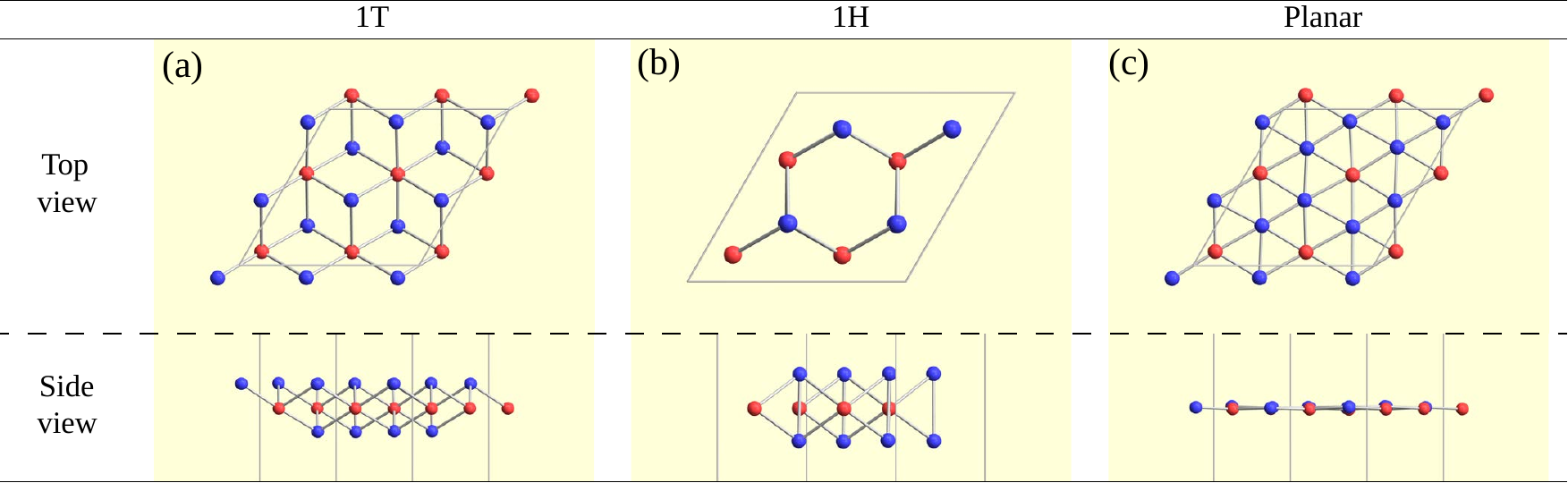}
\caption{Top and side views of initial structures 
for (a) 1T, (b) 1H, and (c) planar.
Red and blue spheres represent atoms A and B, respectively,
where there are 4 atoms of species A and 8 atoms of species B in the unit cell.
}
\label{fig:initial_structure}
\end{figure}

Three types of $2 \times 2$ supercell AB$_2$ structures
(1T , 1H, and planar) as shown in Fig.~\ref{fig:initial_structure} were prepared for each compound as its initial structures.
Here, we note that the name of the 1T structure varies like Trigonal, Octahedral, CdI$_2$-type, $P\bar{3}m1$(164), and D$_{\rm3d}^3$~\cite{Haastrup_2018,Kuc_2015},
while
the name of the 1H structure varies like Hexagonal, Trigonal prismatic, MoS$_2$-type, $P\bar{6}m2$(187), and D$_{\rm3h}^1$ in literatures~\cite{Haastrup_2018,Kuc_2015}.
We made the initial structures include fluctuation from -0.05 \AA\ to 0.05 \AA\ with respect to the ideal structures for atomic coordinates.
The initial spin configurations were prepared as ferromagnetic spin states
whose initial spin density is generated by a superposition of atomic charge density with biased populations of up and down spin charge densities of each atom.
The initial lattice vectors for variable cell optimizations were set to large enough
to allow large structural change from the initial structure
and to avoid trapping of local minima
as far as possible.
The periodic slab approach with vacuum layer of 15 \AA\ 
was used to avoid interaction between periodic layers.

\begin{figure}[htb]
\includegraphics[width=0.95\linewidth]{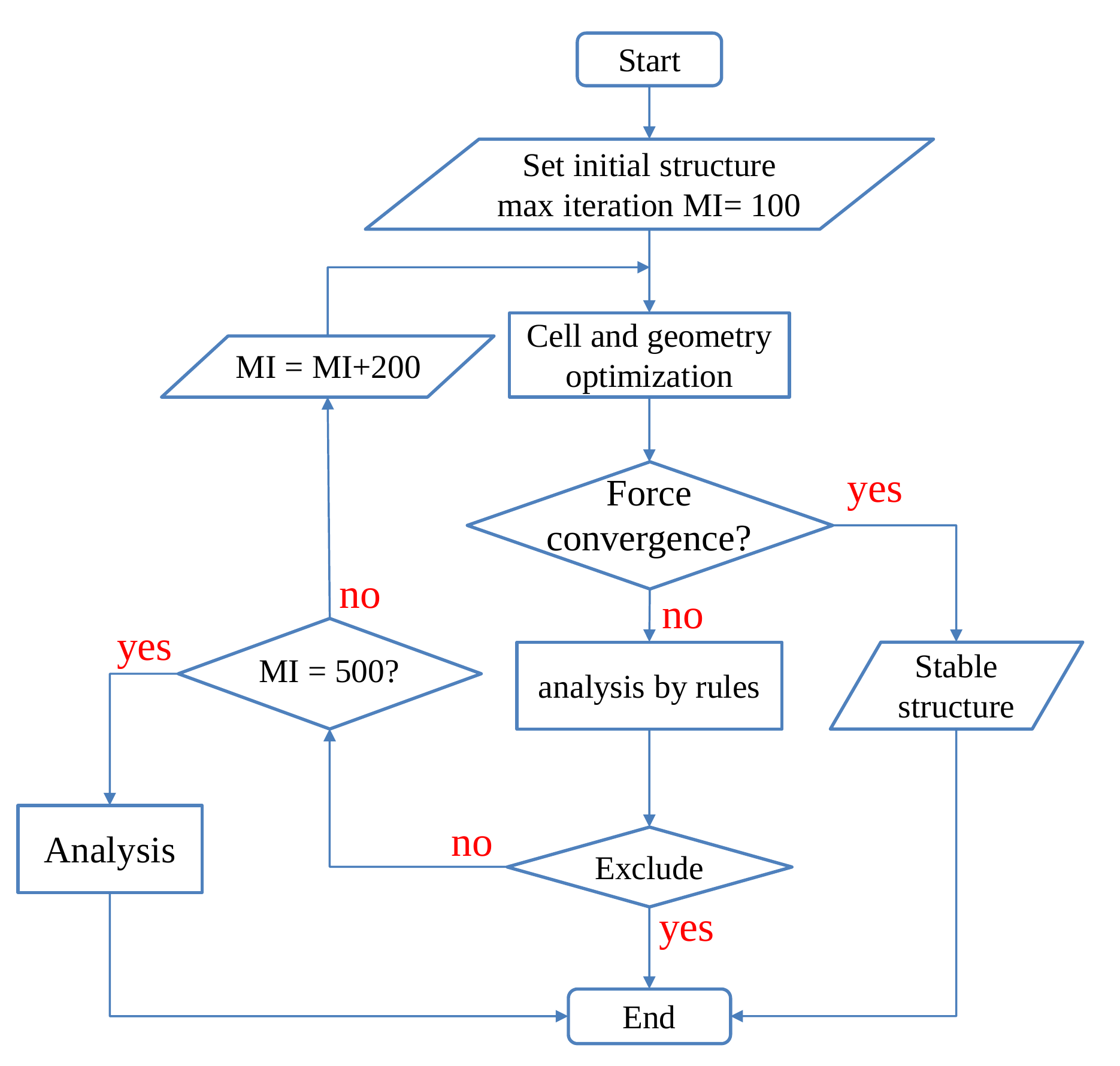}
\caption{The workflow based on the five conditional branch rules used to perform the cell and geometry optimizations for the structure map construction.}
\label{fig:workflow}
\end{figure}

After reaching 100 iterations for the geometry optimization,
we applied the following 5 conditional branch rules
as illustrated in Fig.~\ref{fig:workflow}.

\begin{figure}[htb]
\includegraphics[width=0.95\linewidth]{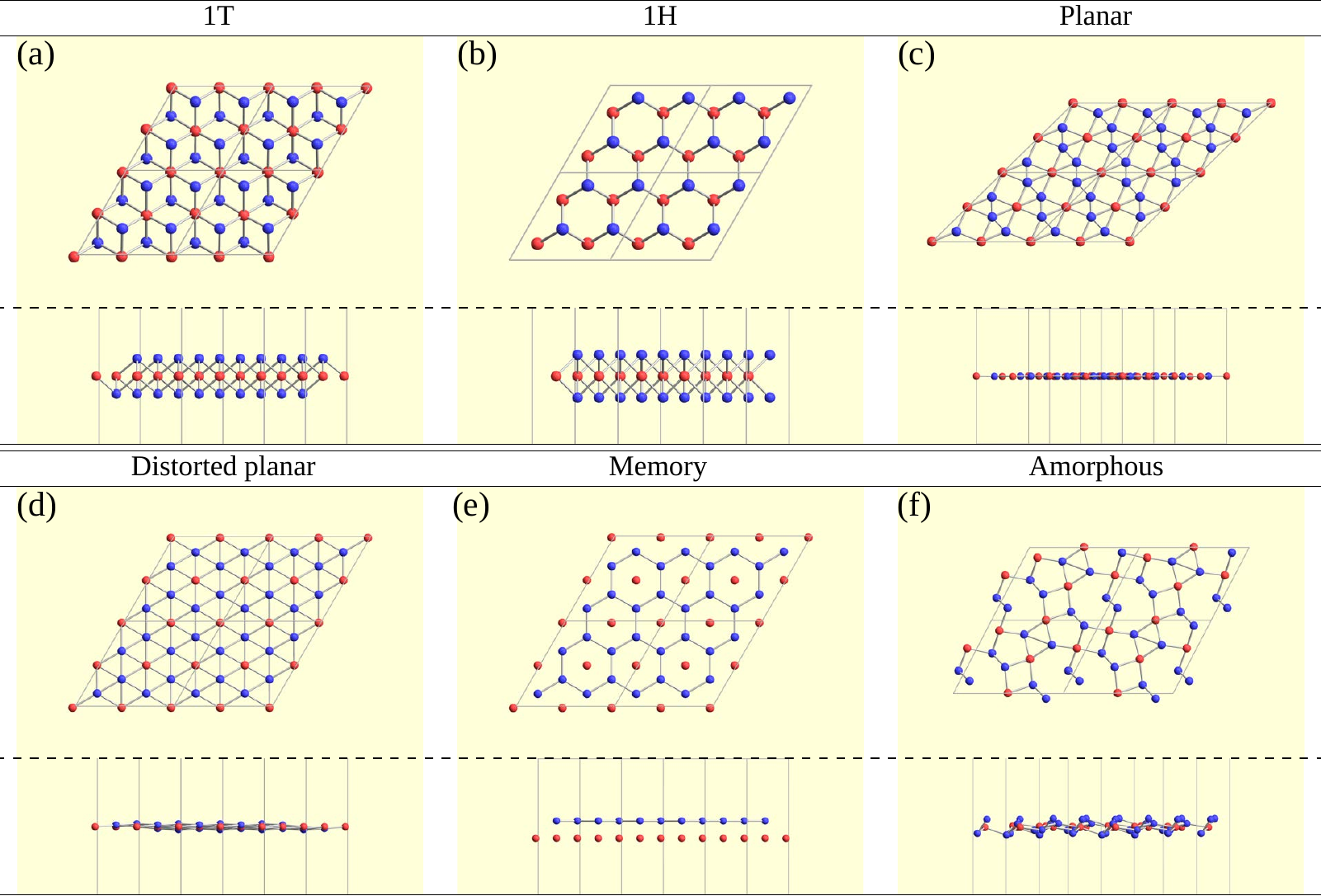}
\caption{Classification of final structures.
Top and side views of examples of final structures 
for (a) 1T, (b) 1H, (c) planar$^{\dagger}$, (d) distorted planar$^{\dagger\dagger}$,
(e) memory$^\S$, and (f) amorphous.
$^{\dagger}$ The difference of the length of each side of the honeycomb is smaller than 0.05 \AA.
In addition, $55^\circ < \angle$ B$_1$A$B_2$ $< 65^\circ$, where B$_1$ and B$_2$ are neighbouring atoms
which are the first neighbouring atoms from the atom A.
The height of each atom from the average height of atoms is smaller than 0.05 \AA.
$^{\dagger\dagger}$ 
The difference of the length of each side of the honeycomb
is smaller than 0.15 \AA.
In addition, the height of each atom from the average height of atoms is smaller than 0.15 \AA.
$^{\S}$ The height of each atom from the average height of atoms is smaller than 0.2 \AA.
In addition, the height of each atom A from the plane formed by atoms B is larger than 0.5 \AA.
}
\label{fig:final_structure}
\end{figure}

\begin{enumerate}
  \item If all the geometry optimizations for three initial structures converge normally, the calculations are stopped, since stable structures have already been obtained.
  \item If at least one of the three calculations does not converge, and however all the final structures are found to be an amorphous structure,
whose definition will be discussed later on, then the calculations are stopped, and the most stable structure is considered to be amorphous.
  \item If only one calculation does not converge, and this final structure is an amorphous structure which has the lowest energy, then the calculations are stopped, and the most stable structure is considered to be amorphous.
  \item If only one calculation converges, and this final structure has the highest energy while structures of the other two calculations are amorphous, then the calculations are stopped, and the most stable structure is considered to be amorphous.
  \item Otherwise, since the most stable structure cannot be determined, more geometry optimizations are performed until it reaches another 200 iterations, and we apply the above rules again. If it has already reached 500 iterations in total, the calculations are stopped, since it is hard to obtain the most stable structure.
\end{enumerate}
Here, we categorized the final structures into planar,
distorted planar, memory, 1$\mathcal{T}$/1$\mathcal{H}$,
and amorphous as shown in Fig.~\ref{fig:final_structure}.
The details of the definitions of these structures such as the tolerance we used are given in the caption of Fig.~\ref{fig:final_structure}.
All the structural figures in this paper are depicted by Crystallica~\cite{Mathematica, Crystallica}.
Here, we briefly explain the definition of the categorized structures.
The group of ``planar'' means that atoms B form almost flat honeycomb structures,
and atoms A are on the center of the honeycombs.
The group of ``distorted planar'' means that the structures are distorted in the plane from the planar structure, while keeping the planarity with a tolerance.
The group of ``memory'' is defined as a group of structures which are similar to the planar,
but atoms A are shifted from the plane formed by atoms B.
As we will discuss the memory structure later on,
the position of the atom A is bistable,
meaning that the atom A is stabilized either above or below the plane formed by the atoms B.
Since the bistability in the position of the atom A might be utilized as binary digits,
we call the buckled structure ``memory structure'', which is expected as a candidate material for a data storage application with an extremely high areal density. 
In the other cases,
judging from whether the top view of the structure is honeycomb-like,
we classify the structure into the group ``1$\mathcal{T}$/1$\mathcal{H}$''
or the group ``amorphous''.
Specifically, if the variation in the distances between the first neighbouring atoms are smaller than 0.15 \AA,
the structures are categorized into the group ``1$\mathcal{T}$/1$\mathcal{H}$'';
otherwise they are categorized into the group ``amorphous''.
Here, we use ``1$\mathcal{T}$/1$\mathcal{H}$'' instead of ``1T/1H'',
because the space-group symmetries of structures in 1$\mathcal{T}$/1$\mathcal{H}$
can be different from the original symmetry of 1T/1H ($P\bar{3}m1$(164) / $P\bar{6}m2$(187)),
and 1$\mathcal{T}$/1$\mathcal{H}$ can include not only 1T/1H but also other characteristic structures.

In this study, we restricted our attention only on constructing
a structure map for almost all the AB$_2$ compounds
by the high-throughput DFT calculations to search possible 2D structures
which are potentially to be 1T, 1H, planar, or memory structures.
The structure map will give us an encompassing perspective
for 2D structures of AB$_2$ type compounds,
and promote further detailed studies such as finite temperature molecular dynamics
and phonon calculations to critically validate structural stability of each newly found compound.



\section{Results} \label{sec:Results}

\subsection{Structure map of AB$_2$ type monolayers} \label{sec:map}
\begin{figure}[htb]
\includegraphics[width=0.95\linewidth]{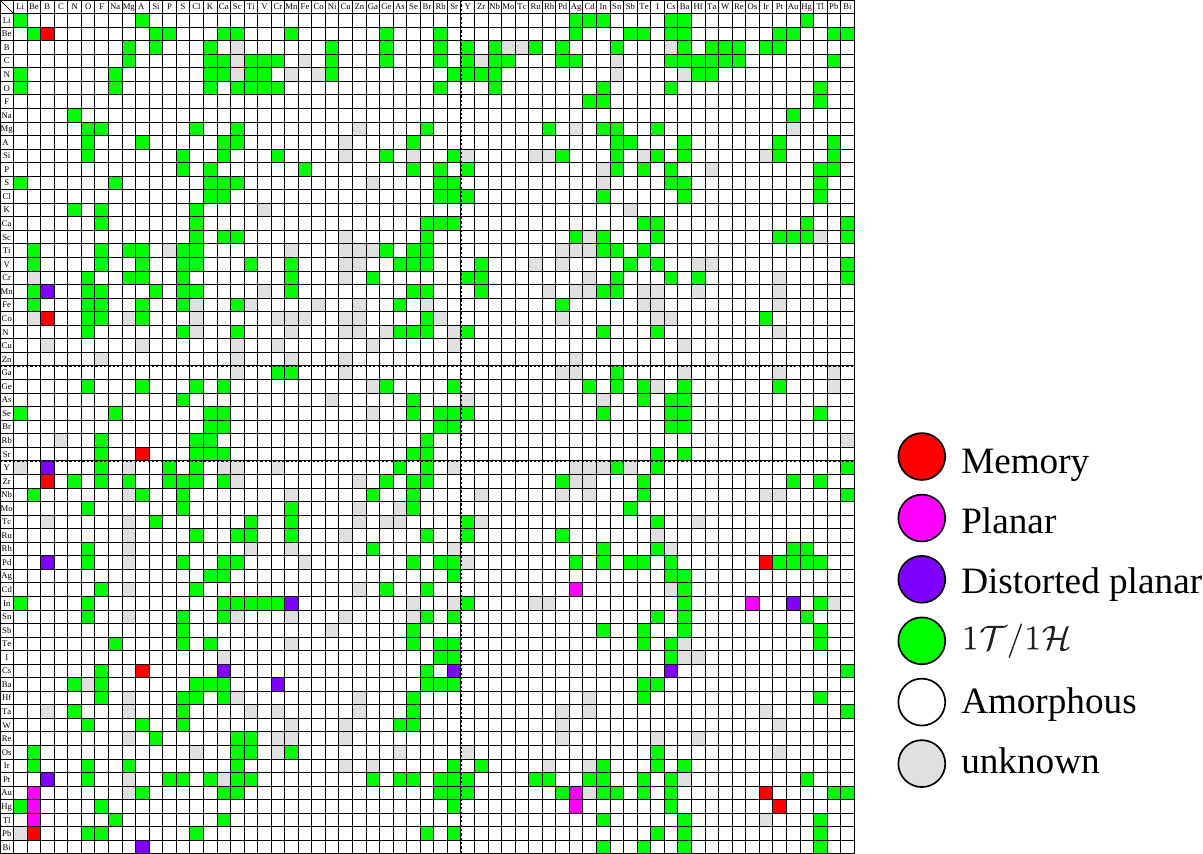}
\caption{Structure map for AB$_2$ type monolayers.
The species of atom A and atom B are specified by the row and column, respectively.
The color of each cell represents the obtained final structure,
whose definitions are given in Fig.~\ref{fig:final_structure}.}
\label{fig:structure_map}
\end{figure}

Based on the classification shown in Fig.~\ref{fig:final_structure},
the most stable converged structures were summarized as a structure map
for AB$_2$ type monolayers in Fig.~\ref{fig:structure_map}.
2332 compounds out of 3844 compounds in total were excluded
as amorphous structures by the conditional branch rules no.~2-4,
leading to remaining 1512 compounds.
Since
among the 1512 compounds
it was hard to obtain the most stable structures for 214 compounds,
whose calculations were terminated after 500 iterations by the conditional branch rule no.~5,
they were classified into the group ``unknown''.
Lastly, we have the screened 1298 compounds,
and call them ``three stable structures'' in this paper,
since all three stable structures for these compounds were obtained by calculations which were performed from three initial structures: 1T, 1H, and planar.
These 1298 compounds were further classified into planar,
distorted planar, memory, 1$\mathcal{T}$/1$\mathcal{H}$,  and amorphous.

From Fig.~\ref{fig:structure_map}, the planar structures include Be, Ag, or Os as the atom B,
while the memory structures include Be, B, Al, Ir, or Pt as the atom B.
However, we could not find any other rules to be the planar or memory structures.
In our DFT calculations, it was difficult to obtain stable structures for
CuX$_2$, XCu$_2$, ZnX$_2$, XZn$_2$, and almost all XM$_2$
(M=Fe, Co, Ni, Cu, Zn, Nb, Mo, Tc, Ru, Rh, Ta, W, Re, Os; X = arbitrary elements) compounds
due to their slow convergence or preference for amorphous structures.
It makes the shape of the distribution of the 1$\mathcal{T}$/1$\mathcal{H}$ structures sparse and striped.

\subsection{Space-group classification} \label{sec:spg}
\begin{figure}[htb]
\center
\includegraphics[width=0.95\linewidth]{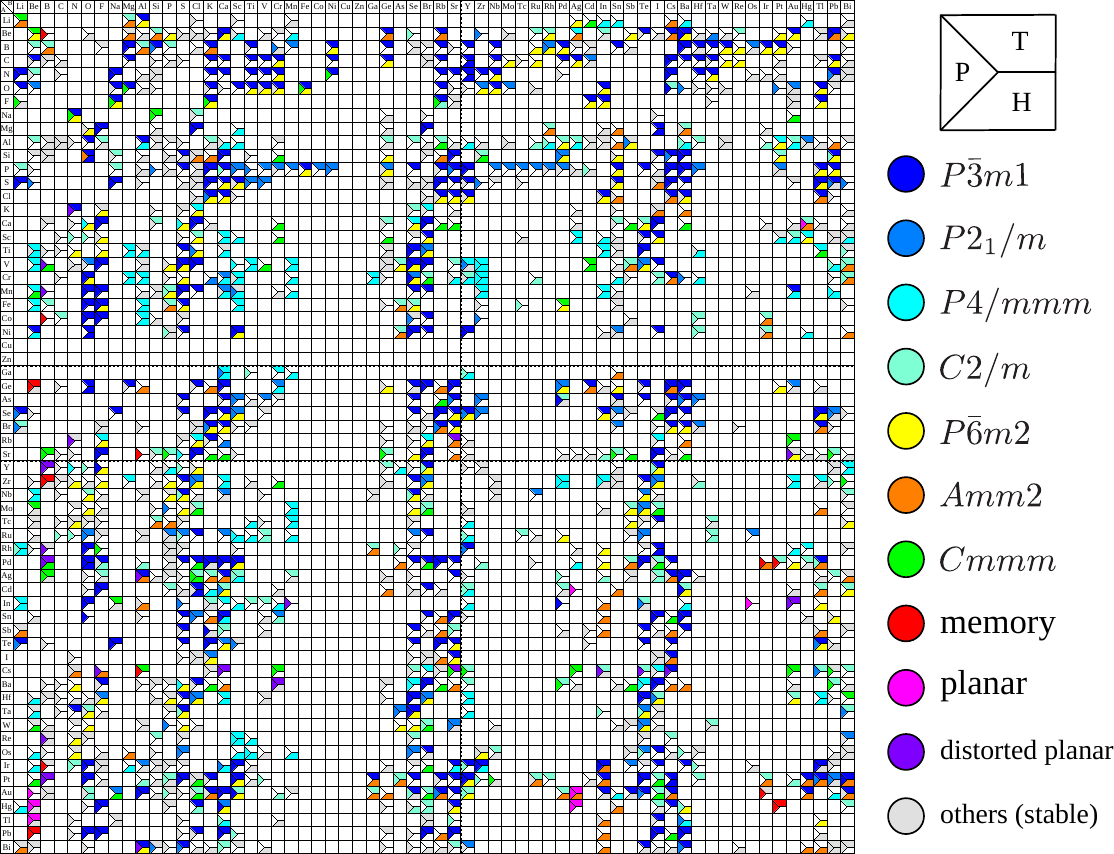}
\caption{Space-group classification. 
The species of atom A and atom B are specified by the row and column, respectively.
The upper, lower and left parts of each cell represent
the initial structures 1T (T), 1H (H), and planar (P) in Fig.~\ref{fig:initial_structure}.
The color of each part of a cell represents the obtained final structure's space-group.
Each cell can have more than one colored parts due to the acceptable energy range 0.03 Hartree ($\approx$ 0.82 eV)
per the unit cell including 12 atoms.
}
\label{fig:space_group}
\end{figure}

Although the rough screening method discussed above is useful to classify the most stable structures intuitively,
it is not sufficient to understand the classification of complicated structures.
For more detailed discussion, 
we classified the structures of 1298 compounds in the group ``three stable structures''
by an analysis based on space-group.
Their space-groups, which were searched by a program code Spglib~\cite{spglib}
with a distance tolerance 0.5 \AA,
are summarized in Fig.~\ref{fig:space_group}.
The structure map is also provided as an interactive website~\cite{AB2_2D_DB}
linked with a customized version of OpenMX Viewer~\cite{OpenMX_Viewer}.
The 1T (e.g. TiS$_2$)  and 1H (e.g. MoS$_2$) phases correspond to
the space-groups $P\bar{3}m1$ (164) and $P\bar{6}m2$ (187), respectively.
Planar and distorted planar structures belong to $P6/mmm$ (191) or $Cmmm$ (65).
In addition,
in the structure map of Fig.~\ref{fig:space_group},
we set an acceptable energy range 0.03 Hartree ($\approx$ 0.82 eV)
per the unit cell including 12 atoms
for comparison among structures obtained from three initial states
to consider the fact that the 2D materials can exist as metastable structures in real experiments due to their fabrication processes and
interactions with substrates.
The acceptable energy range were chosen properly to make the DFT results cover the experimental ones.
This will be mentioned later on Table~\ref{tab:TMDC_DFT} in the following Sec.~\ref{sec:TMDC_TMDO}.
Therefore, the structure map does not show the space-groups of structures
which are excluded due to the above acceptable energy
even if the structures are in the group ``three stable structures''.

In Fig.~\ref{fig:space_group}, roughly speaking,
most of the 1T structures can be found in
TMDCs and TMDOs, metal dihalides,
dialkali-metal materials including groups XIV, XV, XVI (chalcogenides/oxides), and XVII (halides) compounds, 
dialkaline-earth-metal materials including groups XIV, XV, XVI (chalcogenides/oxides), and XVII (halides) compounds, 
transition metal carbides/nitrides (MXenes) and borides/oxides,
and some alloys.
The compounds' name of the 1T ($P\bar{3}m1$) structures are summarized in Appendix~\ref{sec:App_P3m1}.
Similarly,
most of the 1H structures can be found in
mono-metal compounds including diatoms in groups XV, XVI (chalcogenides), and XVII (halides),
dialkali-metal and dialkaline-earth-metal materials including groups XVI (chalcogenides) and XVII (halides) compounds, 
transition metal carbides/nitrides (BiXenes) and borides/oxides,
and some alloys.
The compounds name of the 1H ($P\bar{6}m2$) structures are summarized in Appendix~\ref{sec:App_P6m2}.
The details of the families of 1T/1H structures are discussed in Sec.~\ref{sec:other_1T_1H}.
 
Besides the groups of 1T/1H, many structures in the space-group $P4/mmm$ (123),
which are composed of three square lattice layers,
can be found in Fig.~\ref{fig:space_group} especially in alloys as listed in Appendix~\ref{sec:App_P4mmm}.
Their structures are similar to the AB type structures
such as FeSe~\cite{QYWang_2012} which have 4-fold rotational symmetry.
Although we do not discuss the $P4/mmm$ structure in detail,
the $P4/mmm$ structure might be an interesting family
for the future 2D materials search.

In the following subsections,
we discuss the obtained 1T/1H
(TMDCs and TMDOs, metal dihalides, MXenes/BiXenes, and other 1T/1H),
planar and distorted planar structures, memory structures,
and other characteristic structures in detail.

\subsection{1T/1H structures} \label{sec:1T_1H}

\subsubsection{TMDCs and TMDOs} \label{sec:TMDC_TMDO}

\begin{figure}[htb]
  \includegraphics[width=0.95\linewidth]{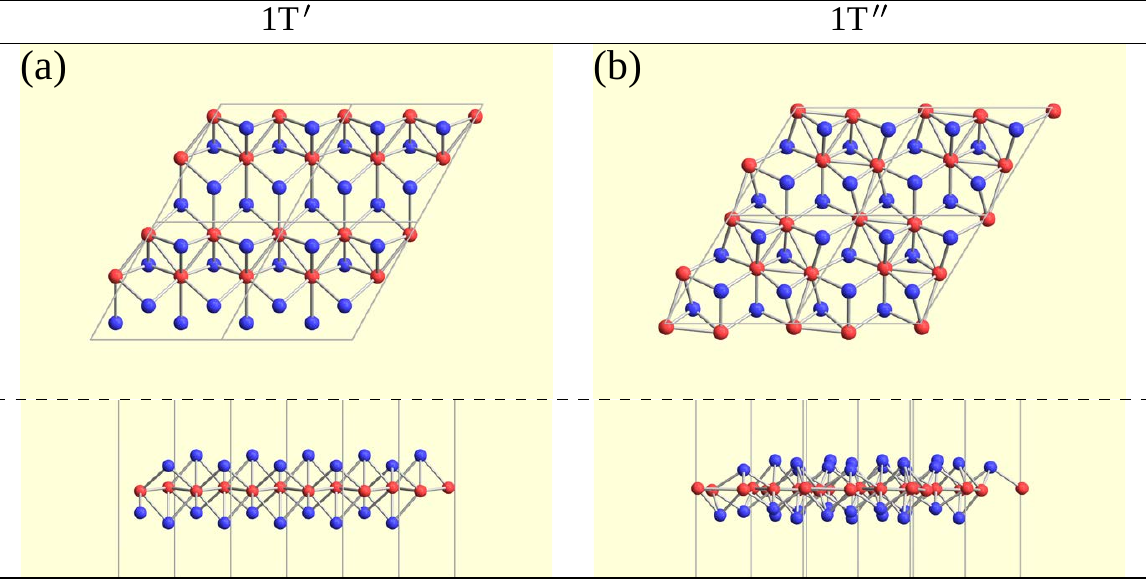}
  \caption{Top and side views of (a) 1T$^\prime$ and (b) 1T$^{\prime \prime}$ structures.}
  \label{tab:TMDC_structures}
\end{figure}

Monolayers of TMDCs are good examples to validate the reliability of our results,
since various kinds of TMDCs have already been synthesized experimentally.
Experimental data reported by J.~Zhou \textit{et al.}~\cite{Zhou_2018} is shown in Table~\ref{tab:TMDC_experiment}.
As described in Ref.~\onlinecite{Zhou_2018} and shown in Fig.~\ref{tab:TMDC_structures},
1T$^\prime$ means the one-dimensional distorted 1T phase in which pairs of metal atoms move closer
to each other perpendicularly,
and 1T$^{\prime \prime}$ means the two-dimensional distorted 1T phase
in which four nearby metal atoms move closer to each other.
While 1T$^\prime$ phase corresponds to $P2_1/m$ (11) in terms of the space-group,
it is difficult to determine the space-group of 1T$^{\prime \prime}$,
since the actual structures are largely distorted and their symmetry is too low.
However, 1T$^{\prime \prime}$ in Ref.~\onlinecite{Zhou_2018} belongs to $P\bar{1}$ or $C2/m$.
It is noted that even though a structure belongs to $P\bar{1}$ or $C2/m$,
it is possible to form a different structure from 1T$^{\prime \prime}$.

Our DFT results corresponding to Table~\ref{tab:TMDC_experiment} are shown in Table~\ref{tab:TMDC_DFT}.
Although our calculations do not include any interactions with a substrate,
our structure table is in good agreement with experimental one.
Other comparisons with previous reported DFT results of TMDCs and TMDOs are summarized
in Tables~\ref{tab:TMDC_TMDO_1}, \ref{tab:TMDC_TMDO_2}, and \ref{tab:TMDC_TMDO_3}.
Most of our results are in good agreement with the other research groups' ones,
though structures which are in magnetic states such as materials including vanadium 
give different stable structures due to the initial spin configuration.
Remember that we chose a ferromagnetic spin state as the initial spin configuration for all the compounds.
The spin magnetic moment for each compound is shown in Appendix~\ref{sec:appendix_magnetic_moment}.
The Tables~\ref{tab:TMDC_TMDO_1}, \ref{tab:TMDC_TMDO_2}, and \ref{tab:TMDC_TMDO_3}
indicate that groups V and VI such as V, Nb, Ta, Cr, Mo, and W tend to form 1H,
while groups IV and X such as Ti, Zr, Hf, Ni, Pd, and Pt tend to form 1T.
This tendency is consistent with 
3D crystal structures of the TMDCs in groups IV, V, VI, VII, and X,
which are known as well-defined layered structures~\cite{Wilson_1969}.

As for TMDOs,
CoO$_2$~\cite{Takada_2003}, MnO$_2$~\cite{Omomo_2003}, and RuO$_2$~\cite{Shikano_2004}
are experimentally known as the 1T intercalant layered structures.
Our results indicate that monolayers of CoO$_2$, MnO$_2$, and RuO$_2$ are 1T, 1T, and 1T$^\prime$, respectively.
This supports that the structures of intercalant layered materials are similar to their monolayers.
Therefore, our analysis on monolayers can contribute to prediction of new intercalant layered structures.

In this subsection, we focused on only transition metals
for discussion of dichalcogenides and dioxides.
However, here we note that alkaline-earth-metals materials
such as BeTe$_2$, MgO$_2$, CaO$_2$, CaS$_2$, CaSe$_2$, CaTe$_2$, SrSe$_2$, and BaS$_2$
also prefer the 1T/1H structure according to the structure map of Fig.~\ref{fig:space_group}.


\begin{table}[htb]
   \center
   \caption{Experimental structures for TMDCs determined by high-resolution STEM imaging reproduced from Ref.~\onlinecite{Zhou_2018}.
   Each color corresponds to the structure symmetry.}
       \label{tab:TMDC_experiment}
   \includegraphics[width=0.95\linewidth]{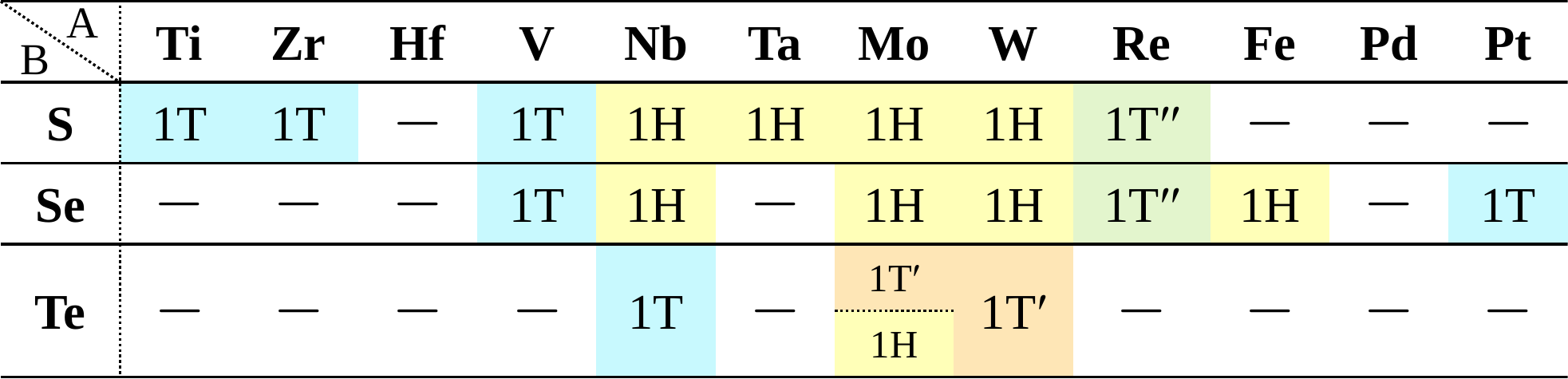}
\end{table}

\begin{table}[htb]
   \center
   \caption{Our DFT results for TMDCs for comparison with Table~\ref{tab:TMDC_experiment}.
   The 1T, 1H, 1T$^\prime$, and 1T$^{\prime\prime}$ correspond to 
   $P\bar{3}m1$, $P\bar{6}m2$, $P2_1/m$ and $P\bar{1}$, respectively.
   Since we allowed the compounds to be multiple stable states
   by the acceptable energy range 0.03 Hartree ($\approx$ 0.82 eV) per the unit cell,
   all the acceptable structures' symmetries are shown in the table.
   The acceptable energy range are chosen properly to make the DFT results cover the experimental ones.}
   \label{tab:TMDC_DFT}
   \includegraphics[width=0.95\linewidth]{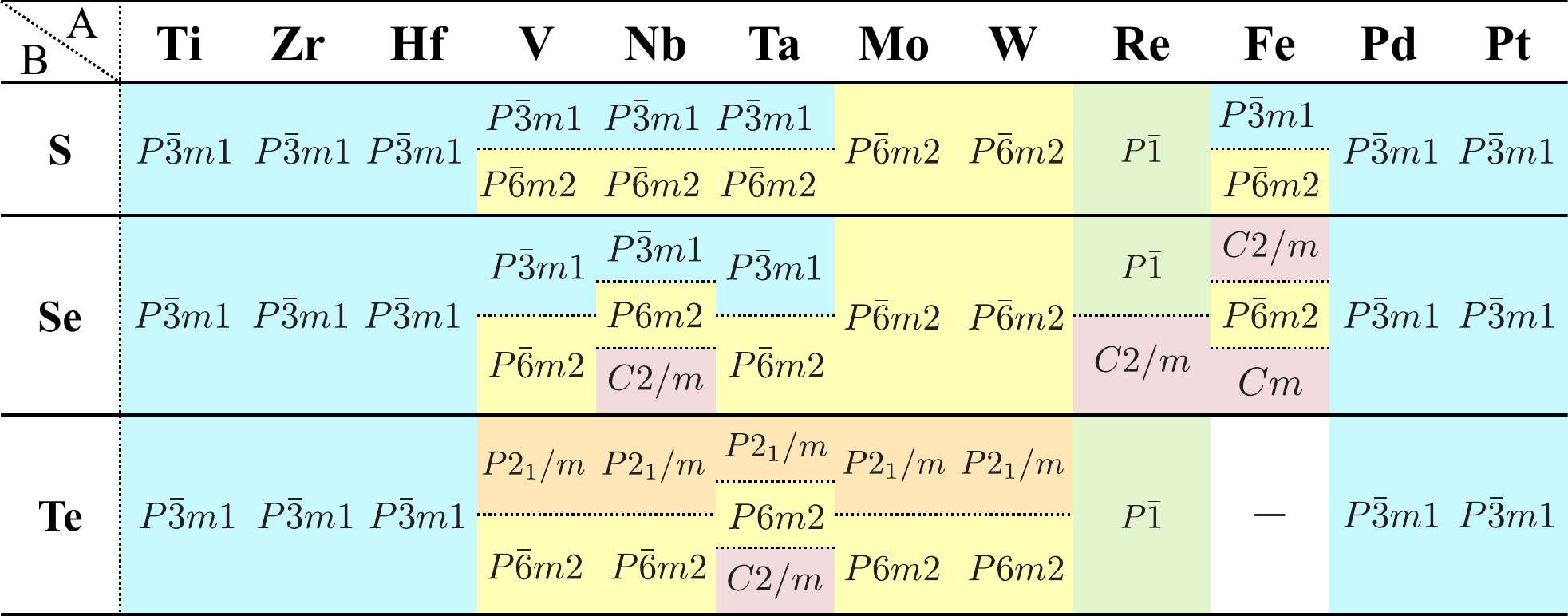}
\end{table}

\begin{table}[htb]
\center
\caption{Comparison with previous reported DFT results of period 4 TMDCs and TMDOs.
        T, H, T$^\prime$, T$^{\prime \prime}$, and M$^\prime$ (see Sec.~\ref{sec:characteristic}) represent the name of stable structures.
        The left upper, right upper, left lower, and right lower parts in each cell
        represent the results in Refs.~\onlinecite{Ataca_2012}, \onlinecite{XZhang_2018}, \onlinecite{Rasmussen_2015}, and our result, respectively.
        Since we allowed the compounds to be multiple stable states
        by the acceptable energy range 0.03 Hartree ($\approx$ 0.82 eV) per the unit cell,
        all the acceptable structures' symmetries are shown in the table.
        The left one is more stable than the right one if two characters appear in the same part of a cell.
        The colors represent the experimentally synthesized compounds summarized in Table~\ref{tab:TMDC_experiment}.}
\includegraphics[width=0.95\linewidth]{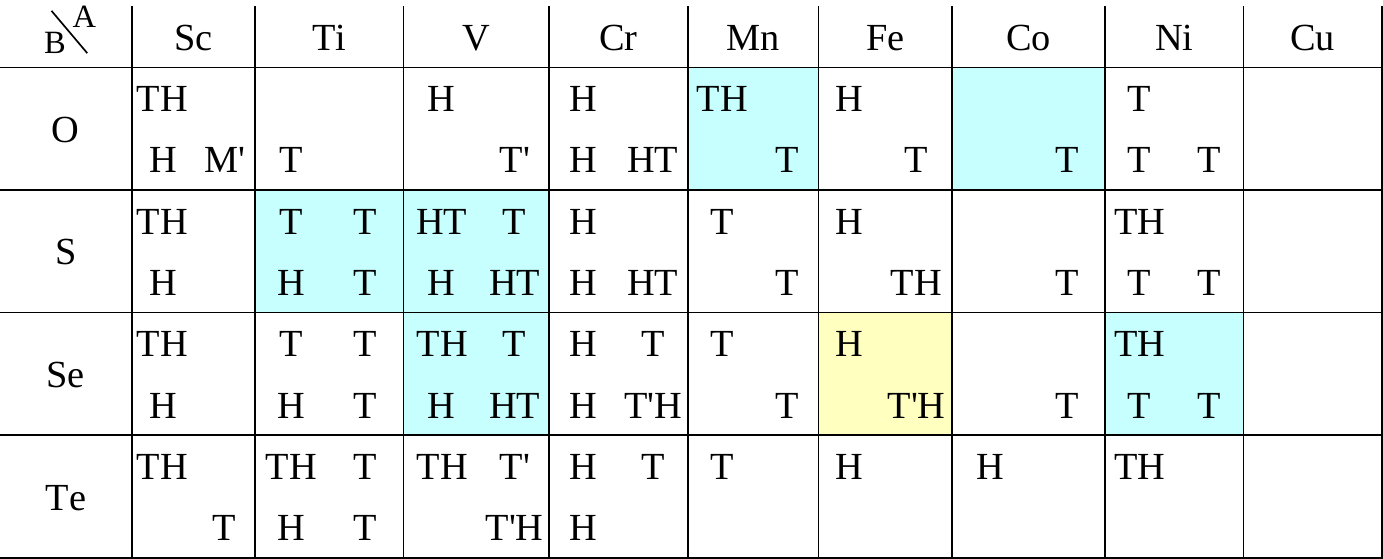}
\label{tab:TMDC_TMDO_1}
\end{table}

\begin{table}[htb]
\center
\caption{Comparison with previous reported DFT results of period 5 TMDCs and TMDOs.
        The left upper, right upper, left lower, and right lower parts in each cell
        represent the results in Refs.~\onlinecite{Ataca_2012}, \onlinecite{XZhang_2018}, \onlinecite{Rasmussen_2015}, and our result, respectively.
}
\includegraphics[width=0.95\linewidth]{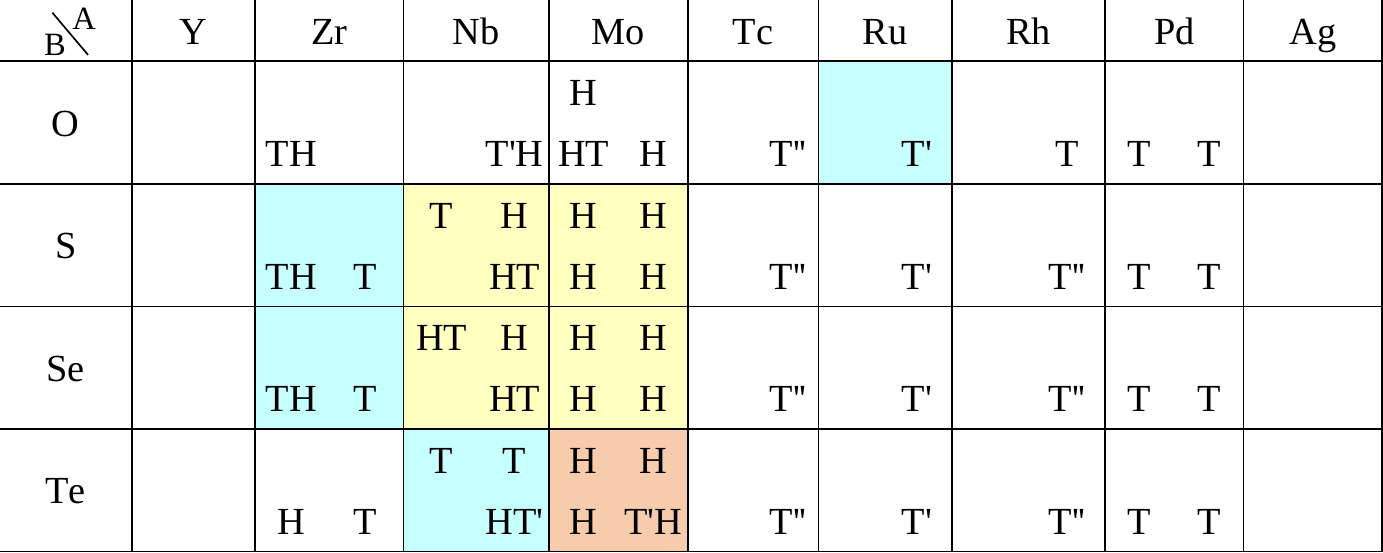}
\label{tab:TMDC_TMDO_2}
\end{table}

\begin{table}[htb]
\center
\caption{Comparison with previous reported DFT results of period 6 TMDCs and TMDOs.
        The left upper, right upper, left lower, and right lower parts in each cell
        represent the results in Refs.~\onlinecite{Ataca_2012}, \onlinecite{XZhang_2018}, \onlinecite{Rasmussen_2015}, and our result, respectively.
}
\includegraphics[width=0.95\linewidth]{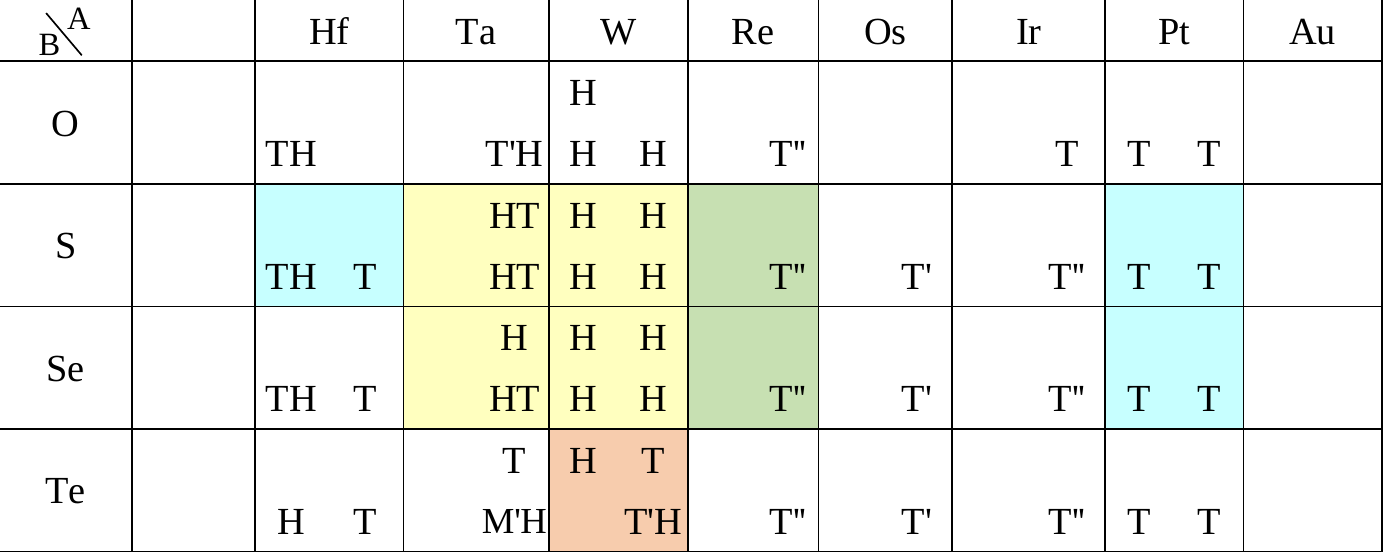}
\label{tab:TMDC_TMDO_3}
\end{table}

\subsubsection{Metal dihalides} \label{sec:dihalides}

A variety of magnetic properties which can be found in
metal dihalides attract great interests~\cite{McGuire_2017,Kulish_2017}, 
since most 2D materials do not possess 
intrinsic ferromagnetism~\cite{Mermin_1966,Huang_2017}.
For example, 1T structures of MX$_2$ (M = V, Cr, Mn, Fe, Co, Ni; X = Cl, Br, I)
have been investigated in Ref.~\onlinecite{Kulish_2017},
and magnetic properties of MX$_2$
(M = alkaline-earth and first row transition metals; X = F, Cl, Br, I)
have been investigated in Ref.~\onlinecite{Lin_2014}.
In Ref.~\onlinecite{YFeng_2018}, the electronic structure, magnetism
and spin transport properties for both 1T and 1H structures of MCl$_2$ (M = V, Cr, Mn, Fe, Co, Ni)
have been investigated.
This magnetic metal dihalides family can also be found in our result of the spin magnetic moment for each compound
shown in Appendix~\ref{sec:appendix_magnetic_moment}.

The most of transition metal dihalides layered structures are known as 
the CdI$_2$ ($P\bar{3}m1$) type or CdCl$_2$ ($R\bar{3}m$) type structure,
whose monolayer have the 1T symmetry (see Ref.~\onlinecite{McGuire_2017} and references therein).
Structures obtained by our calculations are shown in Table~\ref{tab:metal_dihalides}.
Almost all the monolayer transition dihalides prefer the 1T structure.
The families including Y atom prefer the 1H structure
though energetically the 1T structure is allowable,
while ScBr$_2$, ScI$_2$, and FeF$_2$ can also be the 1H structure.

Apart from transition metal dihalides,
Fig.~\ref{fig:space_group} indicates that dihalides including group II, XII or XIV elements
also prefer the 1T structure.
It is noted that some of the dihalides 3D crystal structures including group II or XII elements
such as MgCl$_2$, MgBr$_2$, MgI$_2$, CaI$_2$, and CdBr$_2$
are also known as 
the CdI$_2$ ($P\bar{3}m1$) type or CdCl$_2$ ($R\bar{3}m$) type structure~\cite{Donald_2006,Donald_2009}.




\begin{table}[htb]
\renewcommand{\arraystretch}{1.2}
\newcolumntype{Y}{>{\centering\arraybackslash}p{5mm}} 
\centering
       \caption{Our DFT results for metal dihalides.
               T and H represent the name of stable structures.
               X represent a structure which is neither T nor H.
               The left one is more stable than the right one in each part of a cell.}
\label{tab:metal_dihalides}
\begin{tabular}{c}
    \begin{minipage}{0.45\linewidth}
       \begin{tabular} {c|*{11}{Y}} 
           \backslashbox{B}{A}
            &Be &Mg & &Si &Ge &Sn &Pb & & & & \\
           \hline
           F & &T & & & & &T & & & & \\
           Cl & &T & & &T & &T & & & & \\
           Br & &T & & &T & &T & & & & \\
           I & &T & &T & &T &T & & & & \\
       \end{tabular}
    \end{minipage}
    \hspace{0.04\linewidth}

    \begin{minipage}{0.45\linewidth}
       \begin{tabular} {c|*{11}{Y}} 
           \backslashbox{B}{A}
            &Ca &Sc &Ti &V &Cr &Mn &Fe &Co &Ni &Cu &Zn \\
           \hline
           F &T & & &T & &T &TH &T & & & \\
           Cl &T & & &T & &T & & & & & \\
           Br &T &HT & &T & &T & &T &T & & \\
           I &T &TH & &T & & & & &T & & \\
       \end{tabular}
    \end{minipage}
\end{tabular}
\vspace{4pt}

\begin{tabular}{c}
    \begin{minipage}{0.45\linewidth}
       \begin{tabular} {c|*{11}{Y}} 
           \backslashbox{B}{A}
            &Sr &Y &Zr &Nb &Mo &Tc &Ru &Rh &Pd &Ag &Cd \\
           \hline
           F &T &HT & & & & & & &XT & &T \\
           Cl &T & & & & & &T & &XT & &T \\
           Br &T &HT & & & & &T & &XT & &T \\
           I &T &HT & & & &T & &T & & & \\
       \end{tabular}
    \end{minipage}

    \hspace{0.04\linewidth}
    \begin{minipage}{0.45\linewidth}
       \begin{tabular} {c|*{11}{Y}} 
           \backslashbox{B}{A}
            &Ba & &Hf &Ta &W &Re &Os &Ir &Pt &Au &Hg \\
           \hline
           F &T & & & & & & & & & &T \\
           Cl &T & & & & & & & & & &XT \\
           Br &T & & & & & &XT & & & & \\
           I &T & & & & & &T &T & & & \\
       \end{tabular}
    \end{minipage}
\end{tabular}

\renewcommand{\arraystretch}{1.0}
\end{table}

\subsubsection{MXenes and BiXenes (carbides and nitrides)} \label{sec:MXene}

2D transition metal carbides and nitrides are referred to as MXenes~\cite{Naguib_2014,Tanga_2018,Zhu_2017} or BiXenes~\cite{Sun_2016}.
MXenes and BiXenes are classified by their symmetry;
MXenes have the 1T symmetry and BiXenes have the 1H symmetry~\cite{Sun_2016}. 
Since the MXenes are synthesized in aqueous solution of F$^-$ experimentally,
the surface of MXenes are always terminated with anions such as F$^-$, OH$^-$, and O$^-$~\cite{Wu_2018}.
Therefore,
the multilayered 1T structures
such as Y$_2$C~\cite{Zhang_2014} and Ca$_2$N~\cite{Lee_2013},
and the structures of monolayers
such as Ti$_2$C~\cite{Ying_2017}, Mo$_2$C~\cite{Zhi_2016}, Nb$_2$C~\cite{Naguib_2013}, V$_2$C~\cite{Naguib_2013},
which are 1T structures due to their original crystal structures,
can be slightly different from non-terminated MXenes in our DFT calculations.
Our results and previously reported DFT results~\cite{Chen_2017} are summarized in Table~\ref{tab:MXene}.
Most of our results are in good agreement with previously reported DFT results
except for V$_2$C and V$_2$N, which are in magnetic states (see Appendix~\ref{sec:appendix_magnetic_moment}).
Ti$_2$C, Nb$_2$C, V$_2$C, Y$_2$C, and Ca$_2$N have the 1T symmetry in our results as well as in experimental results, while the 1H symmetry of Mo$_2$C in our result is different from the experimental result~\cite{Zhi_2016}.
However, the DFT result in Ref.~\onlinecite{Chen_2017} has also reported that Mo$_2$C has the 1H symmetry.
Our results predict that
Cr$_2$C, Mo$_2$C, Ru$_2$C, W$_2$C, Re$_2$C,
Cr$_2$N, Nb$_2$N, and Ta$_2$N
can be BiXenes.


\begin{table}[htb]
\renewcommand{\arraystretch}{1.2}
\newcolumntype{Y}{>{\centering\arraybackslash}p{6mm}} 
\center
       \caption{DFT results for MXenes and BiXenes.
               The left tables are DFT results in Ref.~\onlinecite{Chen_2017}.
               The right ones are our results.
               T and H represent the name of stable structures.
               The left one is more stable than the right one in each part of a cell.}
\label{tab:MXene}
\begin{tabular}{c}
    \begin{minipage}{0.45\linewidth}
       \begin{tabular} {c|*8{Y}} 
           \backslashbox{A}{B}     &
           Sc &Ti &V &Cr &Mn &Fe &Co &Ni \\ 
           \hline
           C &T &T &H &T &T & & & \\
           N &H &T &H &H &T & & & \\
       \end{tabular}
    \end{minipage}
    \hspace{0.02\linewidth}

    \begin{minipage}{0.45\linewidth}
       \begin{tabular} {c|*8{Y}} 
           \backslashbox{A}{B}     &
           Sc &Ti &V &Cr &Mn &Fe &Co &Ni \\ 
           \hline
           C & &T &T &TH & & & &T \\
           N & &T &T &TH & & & &T \\
       \end{tabular}
    \end{minipage}
\end{tabular}
\vspace{4pt}

\begin{tabular}{c}
    \begin{minipage}{0.45\linewidth}
       \begin{tabular} {c|*8{Y}} 
           \backslashbox{A}{B}     &
           Y &Zr &Nb &Mo &Tc &Ru &Rh &Pd \\
           \hline
           C & &T &T &H & & & & \\
           N & &T &T &H & & & & \\
       \end{tabular}
    \end{minipage}

    \hspace{0.02\linewidth}
    \begin{minipage}{0.45\linewidth}
       \begin{tabular} {c|*8{Y}} 
           \backslashbox{A}{B}     &
           Y &Zr &Nb &Mo &Tc &Ru &Rh &Pd \\
           \hline
           C &T & &T &H & &H & &T \\
           N &T &T &TH & & & & & \\
       \end{tabular}
    \end{minipage}
\end{tabular}
\vspace{4pt}

\begin{tabular}{c}
    \begin{minipage}{0.45\linewidth}
       \begin{tabular} {c|*8{Y}} 
           \backslashbox{A}{B}     &
             &Hf &Ta &W &Re &Os &Ir &Pt\\
           \hline
           C & &T &T & & & & &         \\
           N & &T &H & & & & &         \\
       \end{tabular}
    \end{minipage}

    \hspace{0.02\linewidth}
    \begin{minipage}{0.45\linewidth}
       \begin{tabular} {c|*8{Y}} 
           \backslashbox{A}{B}     &
             &Hf &Ta &W &Re &Os &Ir &Pt\\
           \hline
           C & &T &T &H &H & & & \\
           N & &T &HT & & & & & \\
       \end{tabular}
    \end{minipage}
\end{tabular}

\renewcommand{\arraystretch}{1.0}
\end{table}

\subsubsection{Other 1T/1H structures} \label{sec:other_1T_1H}

We have already discussed the 1T/1H structures
such as TMDCs, TMDOs, metal dihalides, MXenes, and BiXenes,
in the previous sections.
In order to make families of 1T/1H structures more clearly,
we show the number of 1T/1H structures we obtained for each combination of groups in the periodic table
in Table~\ref{tab:family_1T1H}.
The specific compounds' names of the 1T/1H structures for each combination of groups are listed in Appendix~\ref{sec:family_1T1H}.
For example,
TMDCs and TMDOs in TM-XVI include 39 1T structures and 24 1H structures.
Metal dihalides (I-XVII, II-XVII, and TM-XVII) include 43 1T structures and 25 1H structures.
MXenes and BiXenes are classified in XIV-TM and XV-TM, which include 22 1T structures and 14 1H structures.
Thus, by picking up the large number in Table~\ref{tab:family_1T1H},
we can easily find families of the 1T/1H structures.
In this subsection, we would like to mention other families which can be the 1T/1H structures.

The group TM-XV includes 15 1H structures.
Transition metal dinitrides in TM-XV such as MoN$_2$~\cite{Wang_2015}, ReN$_2$~\cite{Kawamura_2012}, 
OsN$_2$~\cite{Young_2006}, IrN$_2$~\cite{Young_2006,Crowhurst_2006}, and PtN$_2$~\cite{Crowhurst_2006}
are known as multilayered MoS$_2$ type structures (1H).
However, only ReN$_2$ and OsN$_2$ out of them can be the 1H structure in our calculations.
This implies that these multilayers strongly interact with each other in their crystals.
Our calculation also indicates that TiN$_2$, ZrN$_2$, and TcN$_2$ can be the 1H structure.
Here, we note that a different structure type, M-phase, has been reported
for group V transition metal dinitrides (TaN$_2$, NbN$_2$, and VN$_2$) by using DFT calculations~\cite{Wang_2018}.
However, we could not obtain these structures due to the limitation of our choice of initial structures.

According to Ref.~\onlinecite{Hua_2018},
DFT calculations have predicted dialkali-metal monochalcogenides (group XVI) are 1T-phase semiconductors,
which have the inherent layer-by-layer structure with very weak interlayer coupling.
It is found in Table~\ref{tab:family_1T1H} that
not only dialkali-metal monochalcogenides (XVI-I), but also
dialkali-metals and dialkaline-earth-metals materials
including groups XIV, XV, XVI (chalcogenides/oxides), and XVII (halides) compounds,
namely XIV-I, XV-I, XVII-I, XIV-II, XV-II, XVI-II, and XVII-II
such as compounds listed in Appendix~\ref{sec:family_1T}
can be the 1T symmetry.
In addition, 
dialkali-metals halides,
dialkaline-earth-metals chalcogenides/oxides, 
and dialkaline-earth-metals halides,
namely XVII-I, XVI-II, and XVII-II
such as compounds listed in Appendix~\ref{sec:family_1H}
can be the 1H symmetry in our calculations.
Therefore, it would be mentioned that 
these huge families of dialkali-metals and dialkaline-earth-metals materials are
good candidates for new 2D materials
which may have a variety of electronic, magnetic, and optical properties.

Other 1T structures are found in
ditransition metal mono-borides/oxides,
namely XIII-TM and XVI-TM
such as
BTi$_2$, BNi$_2$, BY$_2$, BNb$_2$, BPd$_2$,
BHf$_2$, BTa$_2$, BW$_2$, BIr$_2$, BPt$_2$,
OSc$_2$, OTi$_2$, OV$_2$, OCr$_2$, OFe$_2$,
ONb$_2$,
other ditransition metals such as XIV-TM and XV-TM,
dichalcogenides/dioxides and dihalides such as XIV-XVI, XV-XVI, XVI-XVI, XII-XVII, and XIV-XVII,
alloys,
namely TM-I, TM-II, and TM-TM,
and the other 1T families such as TM-XIII, XVI-XIII, and XVII-XIII,
while the other 1H structures are found in
ditransition metal mono-borides/oxides,
namely XIII-TM and XVI-TM
such as
BNi$_2$, BRu$_2$, BRh$_2$, BPd$_2$, BW$_2$,
BRe$_2$, BIr$_2$, BPt$_2$, OTi$_2$, OV$_2$,
OCr$_2$, OFe$_2$, OZr$_2$, and ONb$_2$,
other ditransition metals such as XIV-TM,
alloys,
namely TM-TM,
and the other 1H families such as TM-XV, XII-II, II-XVI, XVI-XIII and XIV-XIV. 
Again, the specific names of the 1T/1H structures for each combination of groups are listed in Appendix~\ref{sec:family_1T1H}.
As we mentioned in Sec.~\ref{sec:spg},
all the 1T and 1H structures we obtained are listed in Appendices~\ref{sec:App_P3m1} and \ref{sec:App_P6m2}.


\begin{table}[htb]
   \center
   \begin{tabular}{c}
       \begin{minipage}{0.45\linewidth}
          \includegraphics[width=0.95\linewidth]{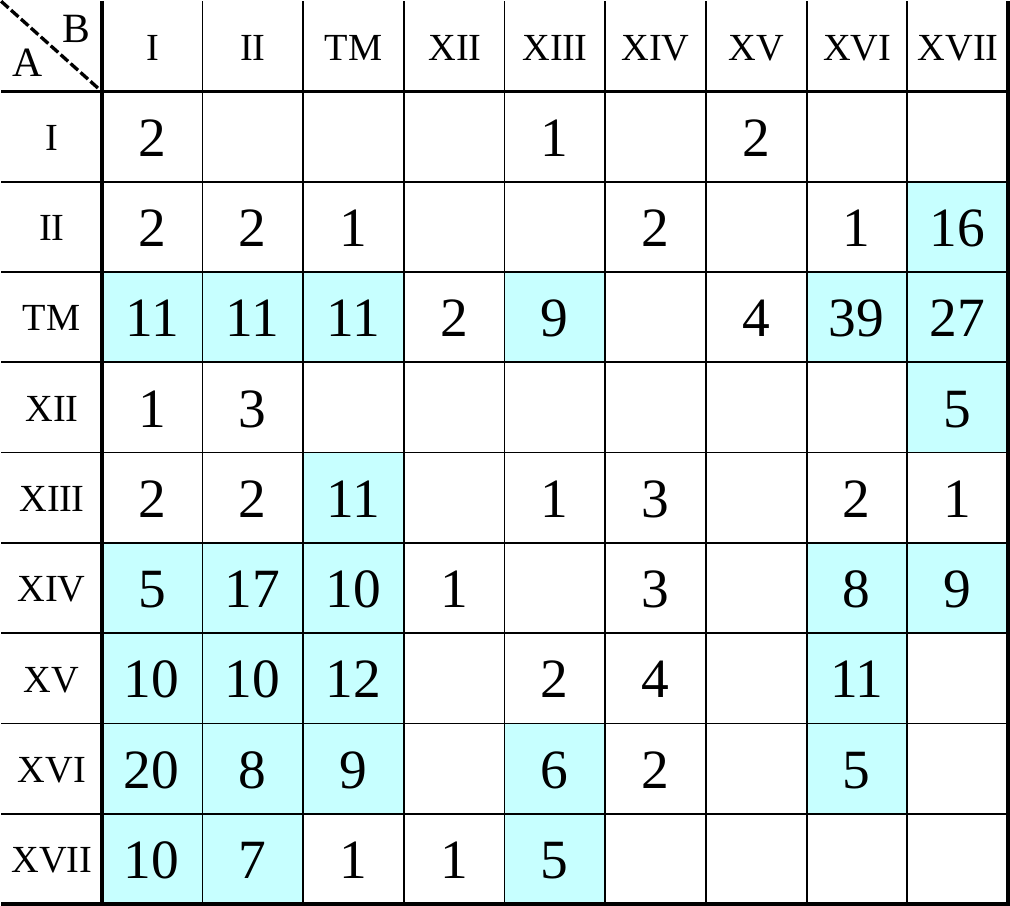}
       \end{minipage}
       \begin{minipage}{0.45\linewidth}
          \includegraphics[width=0.95\linewidth]{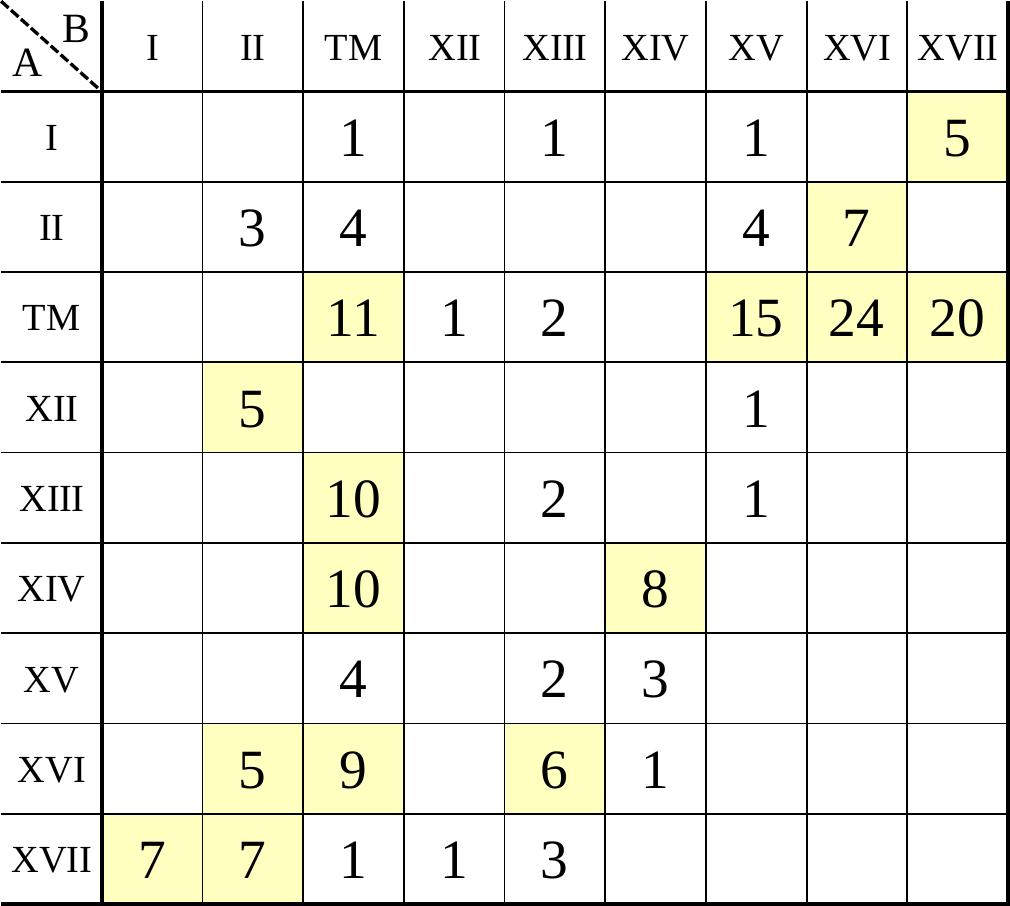}
       \end{minipage}
   \end{tabular}
   \caption{The numbers of compounds of stable 1T (1H) structures classified by groups in the periodic table
           are shown in the left (right) table.
           Cells in which the number is five or more are highlighted to clarify the families of 1T (1H) structures.
   }
   \label{tab:family_1T1H}
\end{table}

\subsection{Planar and distorted planar structures} \label{sec:Planar}
The graphene and h-BN are well known as planar honeycomb monolayers.
As for AB$_2$ type structures,
Cu$_2$S, Cu$_2$Si, and Ag$_2$Bi synthesized on a substrate
are experimentally known as flat structures~\cite{Matsuda_2018}.
However, the stable structures of such compounds could not be obtained in our calculations,
since they are always strongly coupled with a substrate in experiments.
From the structure map of Fig.~\ref{fig:structure_map},
we predicted that 18 compounds can be planar or distorted planar structures
(7 planar structures 
such as CdAg$_2$, InOs$_2$, AuBe$_2$, AuAg$_2$,
HgBe$_2$, HgAg$_2$, and TlBe$_2$,
and 11 distorted planar structures
such as
MnB$_2$, YB$_2$, PdB$_2$, InMn$_2$, InAu$_2$,
CsCa$_2$, CsSr$_2$, CsCs$_2$, BaCr$_2$, PtB$_2$, and
BiAl$_2$).

Assuming that the acceptable energy range 0.03 Hartree per the unit cell as in Fig.~\ref{fig:space_group},
30 compounds can be planar or distorted planar structures
(8 planar structures
such as CaHg$_2$, CdAg$_2$, InOs$_2$, AuBe$_2$, AuAg$_2$, HgBe$_2$, HgAg$_2$, TlBe$_2$,
and 22 distorted planar structures
such as
KN$_2$, VB$_2$, MnB$_2$, RbN$_2$, RbSr$_2$,
SrAu$_2$, YB$_2$, RhB$_2$, PdB$_2$, AgAl$_2$,
InMn$_2$, InAu$_2$, CsF$_2$, CsCa$_2$, CsSr$_2$,
CsIn$_2$, CsTe$_2$, CsCs$_2$, BaCr$_2$, ReB$_2$,
PtB$_2$, BiAl$_2$).
Moreover,
assuming that 
the threshold of the height,
which is the distance between the highest atom and the lowest atom in a direction vertical to the monolayer,
set to be 0.5 \AA, the number of planar-like structures is found to be 74 as listed in Appendix~\ref{sec:App_planar_like}.

\subsection{Memory structures} \label{sec:Memory}
\begin{figure}[htb]
\includegraphics[width=0.98\linewidth]{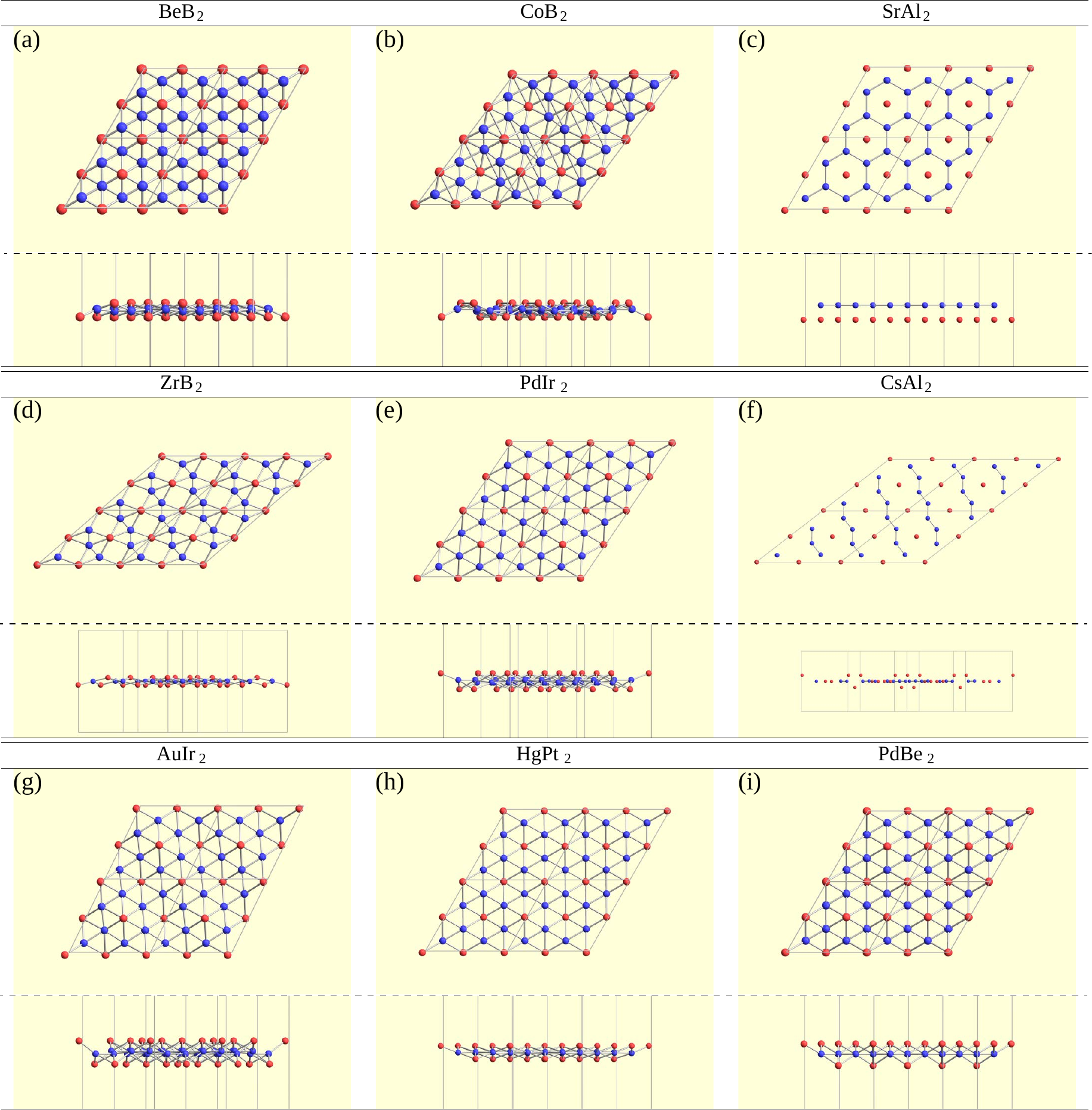}
\caption{Top and side views of 9 memory structures.}\label{fig:memory}
\end{figure}

\begin{figure}[htb]
\includegraphics[width=0.98\linewidth]{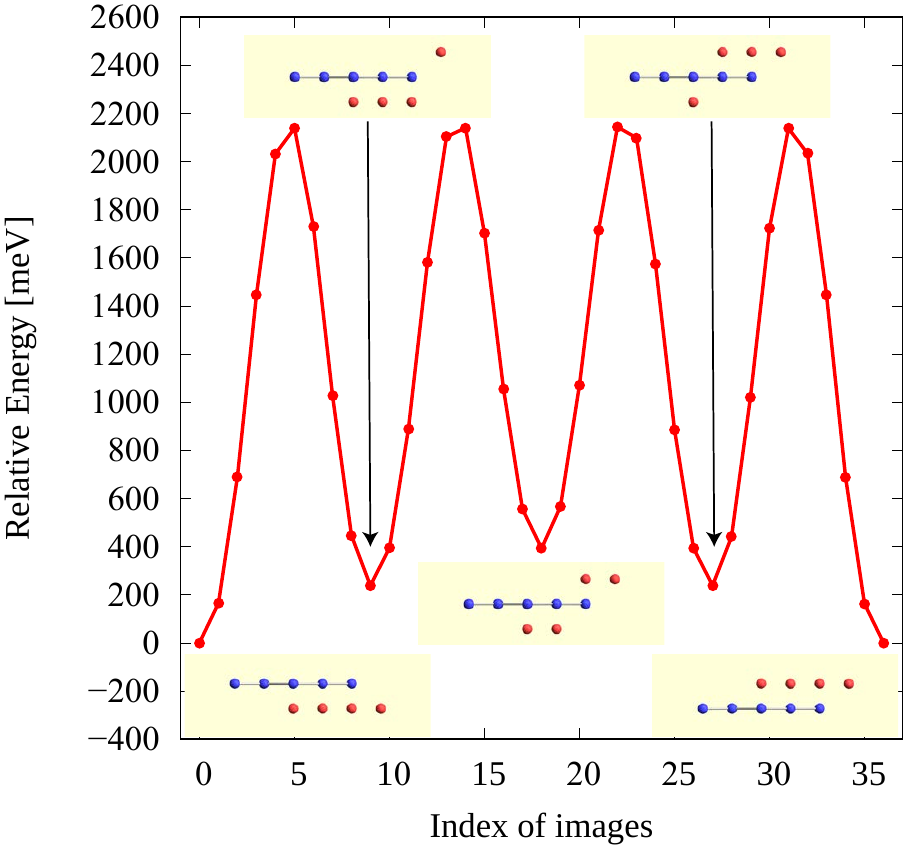}
\caption{The NEB result for the SrAl$_2$ memory structure.
        The five structures on the local minima were prepared to represent a model of a binary digit storage application.
        Each minimum energy path between two of them is obtained by optimizing 8 images on the path.
        The whole energy path is obtained by connecting all the minimum energy paths.
        The corresponding geometrical structures of images on the local energy minima are depicted in the figure.
}\label{fig:NEB_SrAl2}
\end{figure}

In the structure map of Fig.~\ref{fig:structure_map},
we predicted that 9 structures
(AuIr$_2$, BeB$_2$, CoB$_2$, CsAl$_2$, HgPt$_2$,
PbBe$_2$, PdIr$_2$, SrAl$_2$, ZrB$_2$)
can be memory structures.
As we mentioned in Sec.~\ref{sec:Computational_details},
the memory structure is a structure in which
atoms A are above the center of almost flat honeycomb structures formed by atoms B.
Therefore, positions (up or down) of atoms A can represent binary digits.
Since the primitive cell of SrAl$_2$,
whose lattice constant and area are 4.60 \AA\ and 1.83 mm$^2$,
includes only one Sr atom as atom A,
the areal density for the storage application is 5.46$\times 10^{12}$ bit/mm$^2$,
while the areal density of the present HDD storage is about $10^{9}$ bit/mm$^2$~\cite{Gao_2018,Varvaro_2018}.
We performed the nudged elastic band (NEB) calculations\cite{Henkelman_2000} for these memory structures
to confirm whether
it is possible to control each of them
nearly independently with a proper energy barrier
in considering binary digits storage applications.
The NEB calculation for SrAl$_2$ is shown in Fig.~\ref{fig:NEB_SrAl2}.
About 2.2 eV energy barriers of SrAl$_2$ are enough to control it as 
binary digits storage applications at room temperature,
since, for example, the AFM tip can be controlled within the accuracy of less than 1 eV/\AA~\cite{Feng_2018}.
We also confirmed the stability by using the finite temperature molecular dynamics simulation,
which was performed with 1 femtosecond time step for 3 picoseconds at 1000K,
to ensure the structure does not dissociate.
Therefore, it is concluded that SrAl$_2$ is an admirable candidate
for a new binary digits storage application.
The NEB results for the other memory structures 
are shown in Appendix~\ref{sec:appendix_memory_neb}.
Since the AuIr$_2$, PdIr$_2$, and HgPt$_2$ have 0.2-0.8 eV energy barriers,
they are also candidates for memory devices.
The energy barriers are related to the stability of a structure in which atoms A are in the same side.
In addition, the energy barriers are also related to the easiness of an atom A to penetrate the honeycomb hole of atoms B.
Therefore, though BeB$_2$, CsAl$_2$, and PbBe$_2$ belong to memory structure,
their energy barriers are too low
to control structures as a binary digits storage applications
due to their distorted structures and the easiness of penetration.
For example,
since the lattice constants of CsAl$_2$ are much larger than the other memory structures
as the relative size of sphere of atom in the panels in Fig.~\ref{fig:memory} suggests,
the energy barrier is low.
The structures of CoB$_2$ and ZrB$_2$ are too unstable
to obtain NEB results, since the height of protrusive atoms in these materials is much lower
than one of the other memory structures.

Assuming that the threshold of the height,
which is the distance between the highest atom and the lowest atom in a direction vertical to the monolayer,
is set to be 0.5 \AA\ and non-honeycomb structures are allowed
with the acceptable energy range 0.03 Hartree per the unit cell as in Fig.~\ref{fig:space_group},
the number of memory-like structures is found to be 24
as listed in Appendix~\ref{sec:App_memory_like}.


\subsection{Characteristic structures} \label{sec:characteristic}

\begin{figure}[htb]
  \includegraphics[width=0.8\linewidth]{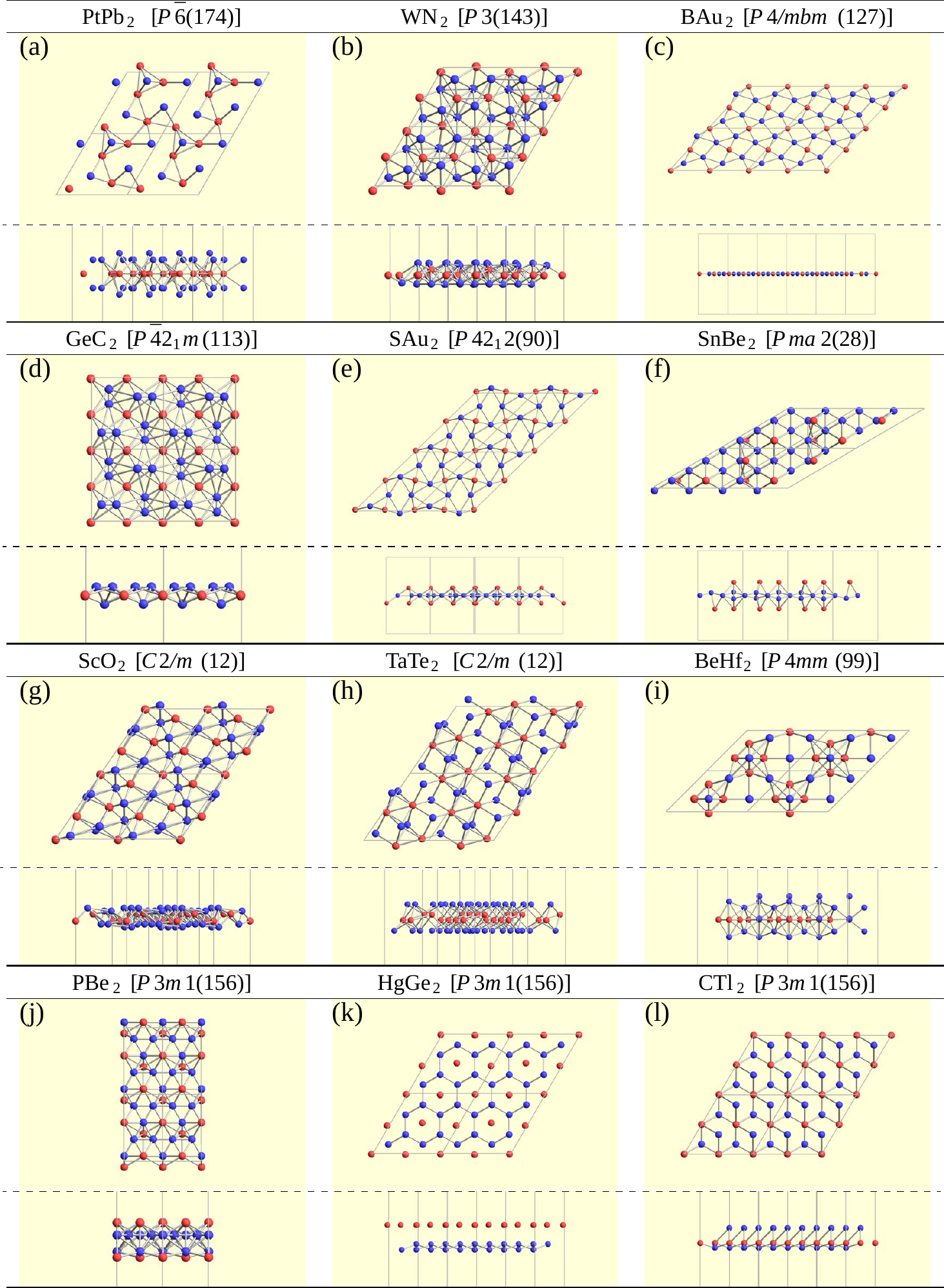}
  \caption{Top and side views of characteristic structures.}
  \label{fig:characteristic_structures}
\end{figure}

Finally, we pick up some characteristic structures which have a minor space-group symmetry.
Their structures are shown in Fig.~\ref{fig:characteristic_structures}.
PtPb$_2$ in $P\bar{6}$ (174) symmetry has 6-fold rotational symmetry
and Pt atoms are in the same plane.
WN$_2$ in $P3$ (143) symmetry is similar to the 1T structure, but the W atoms are buckled.
GeC$_2$ in $P\bar{4}2_1m$ (113) symmetry is a buckled pentagonal planar structure,
which is similar to penta-monolayers in Ref.~\onlinecite{Shao_2018_2} and references therein.
%
As for BAu$_2$ with $P4/mbm$ (127) symmetry and SAu$_2$ with $P42_12$ (90) symmetry,
Au atoms form isogonal distorted square arrangement.
While boron atoms are on the Au plane,
S atoms are above the Au plane similar to the memory structures.
Both of them are pentagonal monolayers.
In SnBe$_2$ in $Pma2$ (28) symmetry, Sn atoms are distributed above and below the plane of buckled trigonal lattice Be monolayer.
Structures of ScO$_2$ in $C2/m$ (12) symmetry and TaTe$_2$ in $C2/m$ (12) symmetry can be viewed as mixtures of 1T and 1H structures
from the top view.
This kind of mixed structure is referred in Ref.~\onlinecite{Wang_2018} as M-phase.
However, the structures of ScO$_2$ and TaTe$_2$ are little different from M-phase
in terms of a boundary structure between 1T and 1H.
Therefore, we called them M$^\prime$-phase in Tables~\ref{tab:TMDC_TMDO_1} and \ref{tab:TMDC_TMDO_3}.
BeHf$_2$ in $P4mm$ (99) symmetry, which looks like a series of baskets, has 4-fold rotational symmetry.
PBe$_2$ in $P3m1$ (156) symmetry is one of the most interesting structures.
The first layer is formed by a buckled honeycomb structure P and Be atoms.
The second layer is a kagome lattice of Be atoms.
The rest of P atoms are above the center of the hexagons of the kagome lattice.
In HgGe$_2$ with $P3m1$ (156) symmetry, Ge atoms form germanene, and Hg atoms are above the center of the hexagons
as memory structures.
CTl$_2$ in $P3m1$(156) symmetry is similar to HgGe$_2$, but C and Tl atoms form a buckled honeycomb structure.

\section{Conclusions} \label{sec:Conclusions}

In this paper,
we have constructed a structure map for AB$_2$ type 2D materials
as shown in Fig.~\ref{fig:structure_map}
on the basis of the high-throughput DFT calculations
by starting from the initial structures: 1T, 1H, and planar structures.
The obtained structures have been classified by their space-group
as shown in Fig.~\ref{fig:space_group}.
The compounds in the major space-groups have been listed in Appendices~\ref{sec:Structure_classification}.
Furthermore, our structure map and database are available on the interactive website~\cite{AB2_2D_DB}
linked with a customized version of OpenMX Viewer~\cite{OpenMX_Viewer},
which enables us to easily visualize obtained structures.

Our results of the well-known families of the 1T/1H ($P\bar{3}m1$/$P\bar{6}m2$) structures such as
\begin{itemize}
   \setlength{\parskip}{0cm}
   \setlength{\itemsep}{0cm}
   \item TMDCs and TMDOs as shown in Tables~\ref{tab:TMDC_experiment}-\ref{tab:TMDC_TMDO_3} in Sec.~\ref{sec:TMDC_TMDO},
   \item metal dihalides as shown in Table~\ref{tab:metal_dihalides} in Sec.~\ref{sec:dihalides},
   \item MXenes and BiXene in Table~\ref{tab:MXene} in Sec.~\ref{sec:MXene},
\end{itemize}
have compared with the experimental data
and previously reported DFT calculations
to confirm reliability of our structure map.

We have summarized the families of the 1T/1H structures as shown in Table~\ref{tab:family_1T1H} in Sec.~\ref{sec:other_1T_1H}.
The left side of Table~\ref{tab:family_1T1H} has predicted that
most of the 1T structures can be found in the following families:
\begin{itemize}
   \setlength{\parskip}{0cm}
   \setlength{\itemsep}{0cm}
   \item TMDCs and TMDOs (TM-XVI),
   \item metal dihalides (II-XVII, and TM-XVII),
   \item other dichalcogenides/dioxides and dihalides such as XIV-XVI, XV-XVI, XVI-XVI, XII-XVII, and XIV-XVII,
   \item MXenes and others in XIV-TM and XV-TM,
   \item ditransition metals such as XIII-TM and XVI-TM,
   \item dialkali-metals such as XIV-I, XV-I, XVI-I, and XVII-I,
   \item dialkaline-earth-metals such as XIV-II, XV-II, XVI-II, and XVII-II,
   \item alloys such as TM-I, TM-II, and TM-TM,
   \item the other 1T families such as TM-XIII, XVI-XIII, and XVII-XIII.
\end{itemize}
The right side of Table~\ref{tab:family_1T1H} has predicted that
most of the 1H structures can be found in the following families:
\begin{itemize}
   \setlength{\parskip}{0cm}
   \setlength{\itemsep}{0cm}
   \item TMDCs and TMDOs (TM-XVI),
   \item alkaline-earth-metal dichalcogenides/oxides (II-XVI),
   \item metal dihalides (I-XVII and TM-XVII),
   \item BiXenes and others in XIV-TM,
   \item ditransition metal mono-borides/oxides and others in XIII-TM and XVI-TM,
   \item dialkali-metals halides, namely XVII-I,
   \item dialkaline-earth-metals chalcogenides and dialkaline-earth-metals halides, namely XVI-II and XVII-II,
   \item other dialkaline-earth-metals such as XII-II,
   \item alloys, namely TM-TM,
   \item the other 1H families such as TM-XV, XVI-XIII, and XIV-XIV. 
\end{itemize}
The specific compounds' names of the 1T/1H structures for each combination of groups have been given in Appendix~\ref{sec:family_1T1H}.

In addition, 
from the structure map of Fig.~\ref{fig:structure_map},
we also predicted the following structures:
\begin{itemize}
   \setlength{\parskip}{0cm}
   \setlength{\itemsep}{0cm}
   \item 7 planar structures such as
CdAg$_2$, InOs$_2$, AuBe$_2$, AuAg$_2$,
HgBe$_2$, HgAg$_2$, and TlBe$_2$,
   \item 11 distorted planar structures such as
MnB$_2$, YB$_2$, PdB$_2$, InMn$_2$, InAu$_2$,
CsCa$_2$, CsSr$_2$, CsCs$_2$, BaCr$_2$, PtB$_2$, and
BiAl$_2$,
   \item 9 memory structures such as AuIr$_2$, BeB$_2$, CoB$_2$, CsAl$_2$, HgPt$_2$,
PbBe$_2$, PdIr$_2$, SrAl$_2$, and ZrB$_2$,
\end{itemize}
Furthermore, the NEB calculations have been performed for memory structures
as shown in Figs.~\ref{fig:NEB_SrAl2} and \ref{fig:NEB_memory}.
The NEB results have supported a possibility to control
the memory structures (SrAl$_2$, AuIr$_2$, PdIr$_2$, and HgPt$_2$) as binary digits data storage applications.

We have also found other characteristic structures as shown in Fig.~\ref{fig:characteristic_structures}:
\begin{itemize}
   \setlength{\parskip}{0cm}
   \setlength{\itemsep}{0cm}
   \item the 6-fold rotational symmetry structure (PtPb$_2$),
   \item pentagonal monolayers (GeC$_2$, BAu$_2$, and SAu$_2$),
   \item buckled trigonal lattice (SnBe$_2$),
   \item mixed structures of the 1T and 1H structures (ScO$_2$ and TaTe$_2$),
   \item the 4-fold rotational symmetry structure (BeHf$_2$),
   \item buckled honeycomb and kagome layers (PBe$_2$),
   \item memory-like structure including germanene (HgGe$_2$) or h-CTl (CTl$_2$).
\end{itemize}
They can be new candidates for AB$_2$ type 2D materials.
As we have mentioned in Sec.~\ref{sec:spg},
the $P4/mmm$ structures listed in Appendix~\ref{sec:App_P4mmm}
are also interesting family of 2D materials.

%
%

We expect that the structure map of AB$_2$ type 2D materials
we presented in the paper will give new viewpoints and directions
to search unknown 2D materials for experimentalists,
and promote further researches to investigate feasibility of newly found compounds
and unveil its physical and chemical properties theoretically.
Further researches to construct structure maps for other compounds
such as AB and AB$_3$ type will be an important future
direction to obtain a more comprehensive understanding
for structural trends of 2D compounds and discover
unknown 2D materials with a wide range of stoichiometric composition ratio. 

\section{Acknowledgement}

M.~F. and T.~O. acknowledge the support of Priority Issue
 (creation of new functional devices and high-performance materials
to support next-generation industries) to be tackled by
using Post `K' Computer, Ministry of Education, Culture,
Sports, Science and Technology, Japan.
J.~Z. acknowledges the financial support of
University of Science and Technology of China.

\appendix

\section{List of basis sets} \label{sec:List_basis}
The basis sets we used in this paper are listed in Table.~\ref{tab:basis_set}.
For example, Fe6.0H-s3p2d1 means that three, two, and one optimized radial functions
were allocated for the $s$, $p$, and $d$ orbitals, respectively for Fe atoms with the ``hard'' pseudopotential,
and the cutoff radius of 6 Bohr was chosen.

\begin{table}[htb]
\centering
  \caption{List of basis sets}
\begin{tabular}{c}
\centering
    \begin{minipage}{0.95\linewidth}
    \begin{tabular}{l@{\hspace{12pt}}l@{\hspace{12pt}}l}
    \hline
    Li8.0-s3p2    &  Fe6.0H-s3p2d1  &  Cd7.0-s3p2d2    \\
    Be7.0-s3p2    &  Co6.0H-s3p2d1  &  In7.0-s3p2d2    \\
    B7.0-s2p2d1   &  Ni6.0H-s3p2d1  &  Sn7.0-s3p2d2    \\
    C6.0-s2p2d1   &  Cu6.0H-s3p2d1  &  Sb7.0-s3p2d2    \\
    N6.0-s2p2d1   &  Zn6.0H-s3p2d1  &  Te7.0-s3p2d2f1  \\
    O6.0-s2p2d1   &  Ga7.0-s3p2d2   &  I7.0-s3p2d2f1   \\
    F6.0-s2p2d1   &  Ge7.0-s3p2d2   &  Cs12.0-s3p2d2   \\
    Na9.0-s3p2d1  &  As7.0-s3p2d2   &  Ba10.0-s3p2d2   \\
    Mg9.0-s3p2d2  &  Se7.0-s3p2d2   &  Hf9.0-s3p2d2    \\
    Al7.0-s2p2d1  &  Br7.0-s3p2d2   &  Ta7.0-s3p2d2    \\
    Si7.0-s2p2d1  &  Rb11.0-s3p2d2  &  W7.0-s3p2d2     \\
    P7.0-s2p2d1f1 &  Sr10.0-s3p2d2  &  Re7.0-s3p2d2    \\
    S7.0-s2p2d1f1 &  Y10.0-s3p2d2   &  Os7.0-s3p2d2    \\
    Cl7.0-s2p2d1f1&  Zr7.0-s3p2d2   &  Ir7.0-s3p2d2    \\
    K10.0-s3p2d1  &  Nb7.0-s3p2d2   &  Pt7.0-s3p2d2    \\
    Ca9.0-s3p2d1  &  Mo7.0-s3p2d2   &  Au7.0-s3p2d2    \\
    Sc9.0-s3p2d1  &  Tc7.0-s3p2d2   &  Hg8.0-s3p2d2f1  \\
    Ti7.0-s3p2d1  &  Ru7.0-s3p2d2   &  Tl8.0-s3p2d2f1  \\
    V6.0-s3p2d1   &  Rh7.0-s3p2d2   &  Pb8.0-s3p2d2f1  \\
    Cr6.0-s3p2d1  &  Pd7.0-s3p2d2   &  Bi8.0-s3p2d2f1  \\
    Mn6.0-s3p2d1  &  Ag7.0-s3p2d2   &                  \\
    \hline
  \end{tabular}
  \end{minipage}
  \end{tabular}
  \label{tab:basis_set}
\end{table}

\section{Structure classification} \label{sec:Structure_classification}

Compounds for each symmetry shown in Fig.~\ref{fig:space_group} in Sec.~\ref{sec:spg}
are listed in the following subsections.

\subsection{$P\bar{3}m1(164)$} \label{sec:App_P3m1}
LiAl$_2$, BeCa$_2$, BeGe$_2$, BeRb$_2$, BeTe$_2$, 
BeCs$_2$, BeBa$_2$, BeAu$_2$, BePb$_2$, BMg$_2$, 
BSi$_2$, BK$_2$, BTi$_2$, BNi$_2$, BGe$_2$, 
BRb$_2$, BY$_2$, BNb$_2$, BPd$_2$, BSn$_2$, 
BBa$_2$, BHf$_2$, BTa$_2$, BW$_2$, BIr$_2$, 
BPt$_2$, CBe$_2$, CMg$_2$, CK$_2$, CCa$_2$, 
CTi$_2$, CV$_2$, CCr$_2$, CNi$_2$, CGe$_2$, 
CRb$_2$, CY$_2$, CNb$_2$, CPd$_2$, CAg$_2$, 
CCs$_2$, CBa$_2$, CHf$_2$, CTa$_2$, CPb$_2$, 
NLi$_2$, NNa$_2$, NK$_2$, NCa$_2$, NTi$_2$, 
NV$_2$, NCr$_2$, NNi$_2$, NRb$_2$, NSr$_2$, 
NY$_2$, NZr$_2$, NNb$_2$, NCs$_2$, NHf$_2$, 
NTa$_2$, NPb$_2$, OLi$_2$, ONa$_2$, OMg$_2$, 
OK$_2$, OSc$_2$, OTi$_2$, OV$_2$, OCr$_2$, 
OFe$_2$, ORb$_2$, ONb$_2$, OIn$_2$, OCs$_2$, 
OTl$_2$, FNa$_2$, FK$_2$, FRb$_2$, FCd$_2$, 
FIn$_2$, FTl$_2$, NaN$_2$, MgF$_2$, MgCl$_2$, 
MgBr$_2$, MgI$_2$, AlO$_2$, AlAl$_2$, SiO$_2$, 
SiS$_2$, SiCa$_2$, SiSr$_2$, SiI$_2$, SiCs$_2$, 
SiBa$_2$, SiPb$_2$, PS$_2$, PK$_2$, PCa$_2$, 
PFe$_2$, PSe$_2$, PRb$_2$, PSr$_2$, PY$_2$, 
PSn$_2$, PTe$_2$, PCs$_2$, PBa$_2$, PTl$_2$, 
PPb$_2$, SLi$_2$, SNa$_2$, SK$_2$, SCa$_2$, 
SSc$_2$, SRb$_2$, SSr$_2$, SY$_2$, SSn$_2$, 
SCs$_2$, SBa$_2$, STl$_2$, SPb$_2$, ClK$_2$, 
ClCa$_2$, ClRb$_2$, ClSr$_2$, ClY$_2$, ClIn$_2$, 
ClCs$_2$, ClBa$_2$, ClTl$_2$, KN$_2$, CaF$_2$, 
CaCl$_2$, CaBr$_2$, CaI$_2$, ScBr$_2$, ScTe$_2$, 
ScI$_2$, TiS$_2$, TiSe$_2$, TiTe$_2$, VF$_2$, 
VS$_2$, VCl$_2$, VSe$_2$, VBr$_2$, VI$_2$, 
CrO$_2$, CrSe$_2$, CrCs$_2$, MnO$_2$, MnF$_2$, 
MnS$_2$, MnCl$_2$, MnSe$_2$, MnBr$_2$, MnAu$_2$, 
FeO$_2$, FeF$_2$, FeS$_2$, CoO$_2$, CoF$_2$, 
CoBr$_2$, NiBe$_2$, NiO$_2$, NiS$_2$, NiSc$_2$, 
NiSe$_2$, NiBr$_2$, NiY$_2$, NiI$_2$, GeO$_2$, 
GeMg$_2$, GeCl$_2$, GeCa$_2$, GeSe$_2$, GeBr$_2$, 
GeSr$_2$, GeCd$_2$, GeTe$_2$, GeCs$_2$, GeBa$_2$, 
AsS$_2$, AsCa$_2$, AsSe$_2$, AsSr$_2$, AsPd$_2$, 
AsSn$_2$, AsTe$_2$, AsCs$_2$, AsBa$_2$, SeLi$_2$, 
SeNa$_2$, SeS$_2$, SeK$_2$, SeCa$_2$, SeSe$_2$, 
SeRb$_2$, SeSr$_2$, SeY$_2$, SeIn$_2$, SeCs$_2$, 
SeBa$_2$, SeTl$_2$, BrK$_2$, BrCa$_2$, BrRb$_2$, 
BrSr$_2$, BrCs$_2$, BrBa$_2$, BrTl$_2$, RbK$_2$, 
SrF$_2$, SrCl$_2$, SrBr$_2$, SrI$_2$, YF$_2$, 
YBr$_2$, YI$_2$, ZrS$_2$, ZrSe$_2$, ZrTe$_2$, 
NbS$_2$, NbSe$_2$, TcI$_2$, RuCl$_2$, RuBr$_2$, 
RhO$_2$, RhY$_2$, RhI$_2$, PdO$_2$, PdF$_2$, 
PdS$_2$, PdCl$_2$, PdK$_2$, PdCa$_2$, PdSc$_2$, 
PdSe$_2$, PdBr$_2$, PdRb$_2$, PdSr$_2$, PdTe$_2$, 
PdCs$_2$, PdTl$_2$, AgAl$_2$, AgCa$_2$, AgSr$_2$, 
AgCs$_2$, AgBa$_2$, CdF$_2$, CdCl$_2$, CdBr$_2$, 
CdBa$_2$, InO$_2$, InPd$_2$, SnO$_2$, SnS$_2$, 
SnCa$_2$, SnBr$_2$, SnSr$_2$, SnI$_2$, SnBa$_2$, 
SbS$_2$, SbK$_2$, SbSe$_2$, SbTe$_2$, SbBa$_2$, 
TeLi$_2$, TeNa$_2$, TeS$_2$, TeK$_2$, TeSe$_2$, 
TeRb$_2$, TeSr$_2$, TeTe$_2$, TeCs$_2$, TeTl$_2$, 
ISr$_2$, ICs$_2$, CsCs$_2$, BaF$_2$, BaCl$_2$, 
BaBr$_2$, BaI$_2$, HfS$_2$, HfSe$_2$, HfTe$_2$, 
TaN$_2$, TaS$_2$, TaSe$_2$, OsBr$_2$, OsI$_2$, 
IrO$_2$, IrSc$_2$, IrSr$_2$, IrY$_2$, IrZr$_2$, 
IrIn$_2$, IrI$_2$, IrBa$_2$, IrTl$_2$, PtO$_2$, 
PtS$_2$, PtK$_2$, PtSc$_2$, PtGa$_2$, PtSe$_2$, 
PtRb$_2$, PtSr$_2$, PtY$_2$, PtIn$_2$, PtTe$_2$, 
PtCs$_2$, PtHg$_2$, PtTl$_2$, PtBi$_2$, AuAl$_2$, 
AuK$_2$, AuCa$_2$, AuSc$_2$, AuSe$_2$, AuRb$_2$, 
AuSr$_2$, AuIn$_2$, AuSb$_2$, AuTe$_2$, AuCs$_2$, 
AuHg$_2$, AuBi$_2$, HgF$_2$, HgCl$_2$, HgCa$_2$, 
HgSr$_2$, HgCs$_2$, TlI$_2$, PbO$_2$, PbF$_2$, 
PbCl$_2$, PbCa$_2$, PbBr$_2$, PbSr$_2$, PbI$_2$, 
PbBa$_2$, BiAl$_2$, BiS$_2$, BiTe$_2$, BiBa$_2$.

\subsection{$P2_1/m(11)$} \label{sec:App_P21m}
BeZr$_2$, BeTa$_2$, BSi$_2$, BCr$_2$, BMn$_2$, 
BRh$_2$, BOs$_2$, CCd$_2$, NPb$_2$, OBe$_2$, 
OZr$_2$, OMo$_2$, OBa$_2$, OPt$_2$, OAu$_2$, 
AlSb$_2$, AlHg$_2$, SiPd$_2$, SiCs$_2$, PSi$_2$, 
PSc$_2$, PTi$_2$, PV$_2$, PCr$_2$, PMn$_2$, 
PCo$_2$, PNi$_2$, PGe$_2$, PSr$_2$, PNb$_2$, 
PMo$_2$, PTc$_2$, PRu$_2$, PRh$_2$, PPd$_2$, 
PHf$_2$, PPt$_2$, SBe$_2$, STi$_2$, SV$_2$, 
SZr$_2$, SPb$_2$, SBi$_2$, ClK$_2$, TiCl$_2$, 
TiBr$_2$, VO$_2$, VY$_2$, VTe$_2$, CrK$_2$, 
CrCa$_2$, CrSe$_2$, CrBa$_2$, MnCa$_2$, MnSc$_2$, 
CoSe$_2$, CoZr$_2$, NiSn$_2$, GaCa$_2$, GeCr$_2$, 
GePd$_2$, GeAu$_2$, AsSc$_2$, AsV$_2$, AsSr$_2$, 
AsZr$_2$, AsBa$_2$, AsHf$_2$, SeLi$_2$, SeSc$_2$, 
SeY$_2$, SeZr$_2$, SeSn$_2$, SePb$_2$, BrLi$_2$, 
BrK$_2$, BrRb$_2$, BrSn$_2$, BrHf$_2$, RbCa$_2$, 
ZrCl$_2$, ZrBr$_2$, ZrI$_2$, NbCa$_2$, NbRu$_2$, 
NbTe$_2$, MoP$_2$, MoTe$_2$, RuO$_2$, RuP$_2$, 
RuS$_2$, RuCa$_2$, RuSe$_2$, RuTe$_2$, RuOs$_2$, 
RhY$_2$, CdCa$_2$, InS$_2$, InTe$_2$, SnCl$_2$, 
SnCr$_2$, TeLi$_2$, TeCa$_2$, TeSc$_2$, TeTl$_2$, 
IK$_2$, CsTl$_2$, HfF$_2$, HfCl$_2$, HfBr$_2$, 
HfI$_2$, TaAs$_2$, TaTe$_2$, TaBi$_2$, WP$_2$, 
WSr$_2$, WTe$_2$, ReBi$_2$, OsS$_2$, OsSe$_2$, 
OsTe$_2$, IrAl$_2$, IrCa$_2$, IrSb$_2$, PtTl$_2$, 
PtPb$_2$, AuNa$_2$, AuRb$_2$, AuTe$_2$, HgCa$_2$, 
TlBa$_2$, BiSi$_2$, BiS$_2$, BiY$_2$, BiBa$_2$.

\subsection{$P4/mmm(123)$} \label{sec:App_P4mmm}
LiIn$_2$, LiSn$_2$, LiBa$_2$, LiHg$_2$, BeSc$_2$, 
BeMn$_2$, BeAg$_2$, MgSc$_2$, AlAl$_2$, AlCa$_2$, 
AlSc$_2$, AlY$_2$, AlRh$_2$, AlAg$_2$, AlBa$_2$, 
AlPt$_2$, AlAu$_2$, KCl$_2$, KBr$_2$, CaO$_2$, 
CaS$_2$, CaSe$_2$, ScCa$_2$, ScSc$_2$, ScAg$_2$, 
ScIn$_2$, ScPt$_2$, ScAu$_2$, ScHg$_2$, ScBi$_2$, 
TiBe$_2$, TiMg$_2$, TiAl$_2$, TiCa$_2$, TiIn$_2$, 
TiHf$_2$, TiTl$_2$, VBe$_2$, VSc$_2$, VTi$_2$, 
VMn$_2$, VY$_2$, VZr$_2$, VPb$_2$, CrMg$_2$, 
CrAl$_2$, CrCa$_2$, CrSc$_2$, CrMn$_2$, CrGa$_2$, 
CrSr$_2$, CrY$_2$, CrZr$_2$, CrSn$_2$, CrHf$_2$, 
MnBe$_2$, MnCa$_2$, MnSc$_2$, MnMn$_2$, MnZr$_2$, 
MnIn$_2$, MnAu$_2$, MnPb$_2$, FeBe$_2$, FeAl$_2$, 
FeSc$_2$, CoAl$_2$, NiBe$_2$, NiHg$_2$, GaCa$_2$, 
GaCr$_2$, GaMn$_2$, GaY$_2$, GeCr$_2$, RbF$_2$, 
RbCl$_2$, RbBr$_2$, RbI$_2$, SrS$_2$, YN$_2$, 
YP$_2$, YAs$_2$, YBi$_2$, ZrMg$_2$, ZrCa$_2$, 
ZrPd$_2$, ZrIn$_2$, ZrAu$_2$, ZrTl$_2$, NbBe$_2$, 
MoMn$_2$, TcTi$_2$, TcMn$_2$, TcY$_2$, RuSc$_2$, 
RuTi$_2$, RuMn$_2$, RuY$_2$, RhLi$_2$, RhAu$_2$, 
RhHg$_2$, PdAg$_2$, AgSc$_2$, CdK$_2$, CdY$_2$, 
InLi$_2$, InCa$_2$, InSc$_2$, InCr$_2$, InY$_2$, 
InBa$_2$, SnCa$_2$, CsBr$_2$, CsI$_2$, BaS$_2$, 
BaSe$_2$, BaTe$_2$, HfBe$_2$, HfCa$_2$, HfY$_2$, 
HfTl$_2$, TaBe$_2$, TaY$_2$, ReSc$_2$, ReTi$_2$, 
OsBe$_2$, OsSc$_2$, OsTi$_2$, OsMn$_2$, IrBe$_2$, 
IrMg$_2$, IrCa$_2$, IrSr$_2$, IrZr$_2$, AuSn$_2$, 
AuPb$_2$, HgLi$_2$, TlCa$_2$, TlBa$_2$.

\subsection{$C2/m(12)$} \label{sec:App_C2m}
LiMg$_2$, BeSi$_2$, BeSe$_2$, BeRu$_2$, BeRh$_2$, 
BeIn$_2$, BeRe$_2$, BBe$_2$, BB$_2$, BAl$_2$, 
BP$_2$, NBe$_2$, OCs$_2$, FRb$_2$, NaCl$_2$, 
MgRh$_2$, MgBa$_2$, AlBe$_2$, AlF$_2$, AlK$_2$, 
AlGe$_2$, AlSe$_2$, AlRb$_2$, AlRu$_2$, AlRh$_2$, 
AlPd$_2$, AlAg$_2$, AlBa$_2$, AlHf$_2$, AlIr$_2$, 
AlAu$_2$, SiNa$_2$, SiZr$_2$, SiPt$_2$, SiPb$_2$, 
PLi$_2$, PNa$_2$, PCa$_2$, PCd$_2$, KSe$_2$, 
CaSi$_2$, CaGe$_2$, CaSn$_2$, CaTe$_2$, CaHg$_2$, 
ScN$_2$, ScO$_2$, ScCl$_2$, ScAu$_2$, TiBi$_2$, 
VAs$_2$, VSr$_2$, VSb$_2$, VPb$_2$, VBi$_2$, 
CrBr$_2$, CrY$_2$, CrRu$_2$, CrI$_2$, CrBi$_2$, 
MnP$_2$, FeB$_2$, FeP$_2$, FeAs$_2$, FeSe$_2$, 
FeTc$_2$, CoC$_2$, CoHf$_2$, CoIr$_2$, NiP$_2$, 
NiAs$_2$, NiHf$_2$, NiIr$_2$, GaTi$_2$, GaY$_2$, 
AsPd$_2$, BrLi$_2$, SrSi$_2$, SrP$_2$, SrGe$_2$, 
SrIn$_2$, SrSn$_2$, SrPb$_2$, YO$_2$, YS$_2$, 
YGe$_2$, YBr$_2$, YI$_2$, ZrHf$_2$, ZrPb$_2$, 
NbO$_2$, NbS$_2$, NbSe$_2$, NbBi$_2$, MoCl$_2$, 
MoBr$_2$, MoSb$_2$, MoI$_2$, TcSi$_2$, TcTa$_2$, 
RuC$_2$, RuN$_2$, RuS$_2$, RuV$_2$, RuY$_2$, 
RuTc$_2$, RuTa$_2$, RhLi$_2$, RhGa$_2$, RhAs$_2$, 
PdF$_2$, PdAl$_2$, PdZr$_2$, PdSb$_2$, PdI$_2$, 
PdPt$_2$, PdHg$_2$, PdTl$_2$, AgF$_2$, AgCl$_2$, 
AgK$_2$, AgSc$_2$, AgMn$_2$, AgPd$_2$, AgIn$_2$, 
AgI$_2$, AgPb$_2$, CdP$_2$, CdCa$_2$, CdY$_2$, 
CdPd$_2$, CdBa$_2$, CdTl$_2$, CdBi$_2$, InK$_2$, 
InTi$_2$, InV$_2$, InTe$_2$, SnCr$_2$, SbCa$_2$, 
SbSr$_2$, CsSe$_2$, CsY$_2$, CsIn$_2$, CsTe$_2$, 
CsTl$_2$, CsBi$_2$, BaY$_2$, BaPd$_2$, BaPb$_2$, 
HfI$_2$, HfAu$_2$, TaO$_2$, TaS$_2$, TaIn$_2$, 
TaTe$_2$, TaBi$_2$, WBr$_2$, WTa$_2$, ReSi$_2$, 
ReSe$_2$, ReCs$_2$, OsNb$_2$, OsTe$_2$, OsBa$_2$, 
IrN$_2$, IrY$_2$, IrHf$_2$, PtAl$_2$, PtP$_2$, 
PtV$_2$, PtAs$_2$, PtZr$_2$, PtRu$_2$, PtRh$_2$, 
PtI$_2$, PtHf$_2$, PtIr$_2$, PtAu$_2$, AuO$_2$, 
AuP$_2$, AuCl$_2$, AuSc$_2$, AuCr$_2$, AuGa$_2$, 
AuSe$_2$, AuBr$_2$, AuY$_2$, AuPd$_2$, AuI$_2$, 
HgCl$_2$, HgSe$_2$, HgBr$_2$, HgI$_2$, HgBa$_2$, 
HgBi$_2$, TlCa$_2$, TlSr$_2$, PbSc$_2$, BiCa$_2$, 
BiSr$_2$.

\subsection{$P\bar{6}m2(187)$} \label{sec:App_P6m2}
LiAl$_2$, LiAg$_2$, BeBe$_2$, BeP$_2$, BeCa$_2$, 
BeRh$_2$, BeSb$_2$, BeBa$_2$, BePt$_2$, BeAu$_2$, 
BeBi$_2$, BNi$_2$, BRu$_2$, BRh$_2$, BPd$_2$, 
BW$_2$, BRe$_2$, BIr$_2$, BPt$_2$, CCr$_2$, 
CGe$_2$, CMo$_2$, CRu$_2$, CAg$_2$, CW$_2$, 
CRe$_2$, CPb$_2$, NCr$_2$, NNb$_2$, NTa$_2$, 
OTi$_2$, OV$_2$, OCr$_2$, OFe$_2$, OZr$_2$, 
ONb$_2$, OIn$_2$, OTl$_2$, FNa$_2$, FK$_2$, 
FCd$_2$, FIn$_2$, FTl$_2$, NaN$_2$, MgO$_2$, 
AlPd$_2$, AlSb$_2$, AlPt$_2$, SiGe$_2$, SiPd$_2$, 
SiSn$_2$, SiPt$_2$, SiPb$_2$, PFe$_2$, PGe$_2$, 
PSn$_2$, PTl$_2$, PPb$_2$, SCa$_2$, SSc$_2$, 
SSn$_2$, STl$_2$, ClK$_2$, ClCa$_2$, ClRb$_2$, 
ClSr$_2$, ClY$_2$, ClIn$_2$, ClBa$_2$, KF$_2$, 
KCl$_2$, CaO$_2$, CaS$_2$, CaSe$_2$, CaTe$_2$, 
ScF$_2$, ScCl$_2$, ScBr$_2$, ScI$_2$, ScHg$_2$, 
TiN$_2$, TiF$_2$, TiCl$_2$, TiBr$_2$, VP$_2$, 
VS$_2$, VAs$_2$, VSe$_2$, VTe$_2$, CrO$_2$, 
CrS$_2$, CrSe$_2$, MnP$_2$, FeF$_2$, FeS$_2$, 
FeSe$_2$, FePd$_2$, NiSc$_2$, GeGe$_2$, GePd$_2$, 
GeSn$_2$, GePt$_2$, SeCa$_2$, SeSc$_2$, SeSr$_2$, 
SeY$_2$, SeIn$_2$, SeTl$_2$, BrK$_2$, BrCa$_2$, 
BrSr$_2$, BrCs$_2$, BrBa$_2$, RbF$_2$, RbCl$_2$, 
RbBr$_2$, SrSe$_2$, SrAu$_2$, YF$_2$, YCl$_2$, 
YBr$_2$, YI$_2$, ZrN$_2$, ZrF$_2$, ZrP$_2$, 
ZrCl$_2$, ZrBr$_2$, ZrI$_2$, NbO$_2$, NbP$_2$, 
NbS$_2$, NbSe$_2$, NbTe$_2$, MoO$_2$, MoS$_2$, 
MoSe$_2$, MoSb$_2$, MoTe$_2$, TcN$_2$, TcSb$_2$, 
TcBi$_2$, RuPd$_2$, RhIn$_2$, PdSc$_2$, PdIn$_2$, 
PdAu$_2$, CdCa$_2$, CdBa$_2$, CdBi$_2$, InTl$_2$, 
SnPb$_2$, TeCa$_2$, TeSr$_2$, TeTl$_2$, IK$_2$, 
ISr$_2$, BaN$_2$, BaS$_2$, HfF$_2$, HfP$_2$, 
HfCl$_2$, HfBr$_2$, HfI$_2$, HfAu$_2$, TaO$_2$, 
TaS$_2$, TaSe$_2$, TaTe$_2$, WO$_2$, WS$_2$, 
WAs$_2$, WSe$_2$, WTe$_2$, ReN$_2$, OsN$_2$, 
OsZr$_2$, PtSc$_2$, PtY$_2$, AuSc$_2$, AuY$_2$, 
HgCa$_2$, HgSr$_2$, HgBa$_2$, TlTl$_2$, BiAl$_2$.

\subsection{$Amm2(38)$} \label{sec:App_Amm2}
LiLi$_2$, LiMg$_2$, LiCs$_2$, BeMg$_2$, BeSi$_2$, 
BeGe$_2$, BeRb$_2$, BeAg$_2$, BeCs$_2$, BePb$_2$, 
BB$_2$, BAl$_2$, BK$_2$, MgRh$_2$, MgSn$_2$, 
MgBa$_2$, AlSn$_2$, AlPb$_2$, SiO$_2$, SiCl$_2$, 
SiK$_2$, SiRb$_2$, SI$_2$, ClSn$_2$, ClCs$_2$, 
ClTl$_2$, KI$_2$, KBa$_2$, CaHg$_2$, VSb$_2$, 
VBi$_2$, CrCs$_2$, CrBi$_2$, FeP$_2$, FeAs$_2$, 
CoIr$_2$, NiAs$_2$, NiIr$_2$, NiBi$_2$, GaSn$_2$, 
GeAl$_2$, GeK$_2$, GeRb$_2$, GeCd$_2$, GeCs$_2$, 
BrRb$_2$, BrTl$_2$, RbSr$_2$, RbBa$_2$, SrSr$_2$, 
MoTl$_2$, TcSi$_2$, TcP$_2$, RhGa$_2$, PdAs$_2$, 
PdSb$_2$, PdIr$_2$, PdTl$_2$, AgAl$_2$, AgK$_2$, 
AgCa$_2$, AgCs$_2$, AgBi$_2$, CdGe$_2$, CdSn$_2$, 
CdTl$_2$, InAl$_2$, InIn$_2$, SnK$_2$, SnRb$_2$, 
SnIn$_2$, SbLi$_2$, SbRb$_2$, SbIn$_2$, SbCs$_2$, 
SbTl$_2$, IRb$_2$, ICs$_2$, CsN$_2$, CsF$_2$, 
BaCa$_2$, BaSr$_2$, BaCs$_2$, BaBa$_2$, OsMg$_2$, 
IrIn$_2$, PtGa$_2$, PtAs$_2$, PtRu$_2$, PtIn$_2$, 
PtIr$_2$, PtHg$_2$, PtTl$_2$, PtBi$_2$, AuK$_2$, 
AuGa$_2$, AuGe$_2$, AuRb$_2$, AuPd$_2$, AuAg$_2$, 
AuCs$_2$, AuHg$_2$, AuBi$_2$, HgSn$_2$, TlIn$_2$, 
PbRb$_2$, PbIn$_2$, BiLi$_2$, BiIn$_2$, BiHg$_2$, 
BiTl$_2$.

\subsection{$Cmmm(65)$} \label{sec:App_Cmmm}
LiLi$_2$, LiCd$_2$, BeBe$_2$, NNi$_2$, OMg$_2$, 
OFe$_2$, FLi$_2$, FNa$_2$, FK$_2$, FRb$_2$, 
NaN$_2$, NaSi$_2$, NaAu$_2$, AlRb$_2$, SiCr$_2$, 
SiZr$_2$, CaCr$_2$, CaRb$_2$, CaSr$_2$, CaSb$_2$, 
CaBa$_2$, ScCr$_2$, ScGe$_2$, TiTl$_2$, VB$_2$, 
VV$_2$, VCd$_2$, CrBr$_2$, MnBe$_2$, MnB$_2$, 
FePd$_2$, AsRb$_2$, SeI$_2$, RbAu$_2$, SrB$_2$, 
SrP$_2$, SrK$_2$, SrCa$_2$, SrGe$_2$, SrSn$_2$, 
SrSb$_2$, SrBa$_2$, SrAu$_2$, SrPb$_2$, YB$_2$, 
ZrCr$_2$, ZrBi$_2$, MoBe$_2$, MoCl$_2$, MoBr$_2$, 
MoI$_2$, RhB$_2$, RhF$_2$, PdB$_2$, PdCl$_2$, 
PdBr$_2$, PdI$_2$, AgB$_2$, AgCl$_2$, InNa$_2$, 
SnCs$_2$, CsF$_2$, CsAl$_2$, CsCa$_2$, CsCr$_2$, 
CsSr$_2$, CsY$_2$, CsAg$_2$, CsIn$_2$, CsTe$_2$, 
CsAu$_2$, CsPb$_2$, BaK$_2$, BaCr$_2$, BaRb$_2$, 
BaPd$_2$, BaSb$_2$, BaPb$_2$, HfBi$_2$, WBe$_2$, 
WBr$_2$, ReB$_2$, IrBr$_2$, PtBe$_2$, PtB$_2$, 
PtCl$_2$, PtNb$_2$, PtI$_2$, AuNa$_2$, AuCl$_2$, 
AuBr$_2$, AuI$_2$, PbCs$_2$.

\subsection{planar} \label{sec:App_planar}
CaHg$_2$, CdAg$_2$, InOs$_2$, AuBe$_2$, AuAg$_2$, HgBe$_2$, HgAg$_2$, TlBe$_2$.

\subsection{distorted planar} \label{sec:App_distorted_planar}
KN$_2$, VB$_2$, MnB$_2$, RbN$_2$, RbSr$_2$,
SrAu$_2$, YB$_2$, RhB$_2$, PdB$_2$, AgAl$_2$,
InMn$_2$, InAu$_2$, CsF$_2$, CsCa$_2$, CsSr$_2$,
CsIn$_2$, CsTe$_2$, CsCs$_2$, BaCr$_2$, ReB$_2$,
PtB$_2$, BiAl$_2$.

\subsection{planar-like structures} \label{sec:App_planar_like}
LiCl$_2$, BAu$_2$, NNi$_2$, OMg$_2$, OFe$_2$, FLi$_2$, FNa$_2$, FK$_2$, FRb$_2$, FAu$_2$,
NaN$_2$, NaSi$_2$, NaBr$_2$, KN$_2$, KBr$_2$, KI$_2$, CaHg$_2$, VB$_2$, MnB$_2$, RbN$_2$,
RbK$_2$, RbSr$_2$, RbI$_2$, RbAu$_2$, SrB$_2$, SrSi$_2$, SrP$_2$, SrGe$_2$, SrSn$_2$, SrAu$_2$,
SrPb$_2$, YB$_2$, RhB$_2$, PdB$_2$, AgB$_2$, AgAl$_2$, AgCl$_2$, AgI$_2$, CdO$_2$, CdAg$_2$,
InMn$_2$, InOs$_2$, InAu$_2$, II$_2$, CsN$_2$, CsF$_2$, CsAl$_2$, CsK$_2$, CsCa$_2$, CsCr$_2$,
CsRb$_2$, CsSr$_2$, CsY$_2$, CsAg$_2$, CsIn$_2$, CsTe$_2$, CsI$_2$, CsCs$_2$, CsAu$_2$, CsPb$_2$,
BaCr$_2$, BaPd$_2$, BaPb$_2$, ReB$_2$, PtB$_2$, AuBe$_2$, AuB$_2$, AuCl$_2$, AuAg$_2$, HgBe$_2$,
HgO$_2$, HgAg$_2$, TlBe$_2$, BiAl$_2$.

\subsection{memory} \label{sec:App_memory}
BeB$_2$, CoB$_2$, SrAl$_2$, ZrB$_2$, PdIr$_2$,
CsAl$_2$, AuIr$_2$, HgPt$_2$, PbBe$_2$.

\subsection{memory-like structures} \label{sec:App_memory_like}
BeB$_2$, OCr$_2$, OPt$_2$, OAu$_2$, SAu$_2$, KSi$_2$, TiB$_2$, CoB$_2$, GeBe$_2$, BrI$_2$,
SrAl$_2$, ZrB$_2$, RhB$_2$, PdIr$_2$, PdPt$_2$, TeB$_2$, CsAl$_2$, BaSi$_2$, BaGe$_2$, HfB$_2$,
IrB$_2$, AuIr$_2$, HgPt$_2$, PbBe$_2$.

\section{Spin magnetic moment map} \label{sec:appendix_magnetic_moment}

\begin{figure}[htb]
\includegraphics[width=0.95\linewidth]{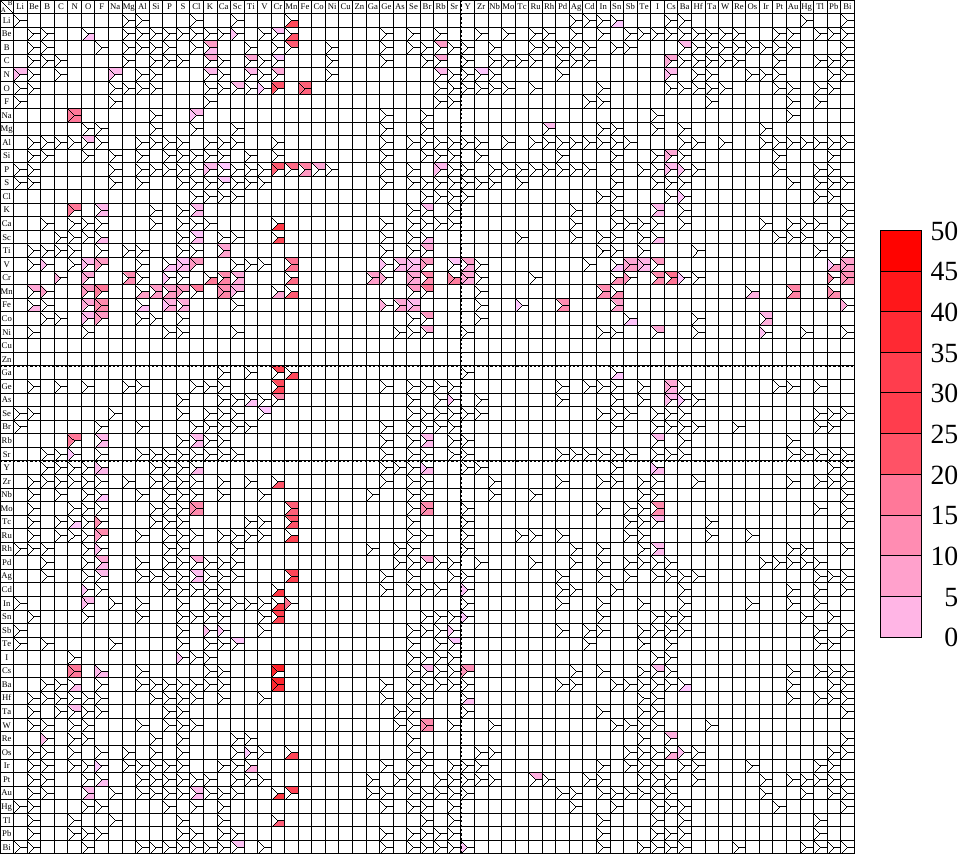}
\caption{Spin magnetic moment map for AB$_2$ type monolayers.
The degree of intensity of red color represents the magnitude of the spin magnetic moment [$\mu_B$/unit cell].}
\label{fig:mag_map}
\end{figure}

The spin magnetic moment per the unit cell including 12 atoms
for each structure is summarized on the structure map of Fig.~\ref{fig:mag_map}.
In our calculations, a ferromagnetic spin state is chosen as the initial spin configuration for all the compounds.
It can be found that 209 compounds have nonzero spin magnetic moment in this figure as listed below.
In particular, compounds including V, Cr, Mn, or Fe as atom A,
or including Cr or Mn as atom B have large spin magnetic moment.
The dihalides, dialkali-metals, and dialkaline-earth-metals might be families of magnetic AB$_2$ monolayers.

LiMn$_2$, BeO$_2$, BeSc$_2$, BeCr$_2$, BeMn$_2$, BK$_2$, BMn$_2$, BRb$_2$, BBa$_2$, CK$_2$,
CTi$_2$, CRb$_2$, CCs$_2$, NLi$_2$, NNa$_2$, NK$_2$, NTi$_2$, NCr$_2$, NRb$_2$, NCs$_2$,
OSc$_2$, OCr$_2$, OFe$_2$, NaN$_2$, NaCl$_2$, MgRh$_2$, AlO$_2$, SiCs$_2$, PK$_2$, PCr$_2$,
PMn$_2$, PFe$_2$, PCo$_2$, PRb$_2$, PCs$_2$, PBa$_2$, SCa$_2$, ClBa$_2$, KN$_2$, KF$_2$,
KCl$_2$, KBr$_2$, KI$_2$, CaCr$_2$, ScF$_2$, ScCl$_2$, ScCr$_2$, ScBr$_2$, ScI$_2$, TiCa$_2$,
TiBr$_2$, VB$_2$, VO$_2$, VF$_2$, VP$_2$, VS$_2$, VCl$_2$, VMn$_2$, VGe$_2$, VAs$_2$,
VSe$_2$, VBr$_2$, VSr$_2$, VY$_2$, VSn$_2$, VSb$_2$, VTe$_2$, VI$_2$, VPb$_2$, VBi$_2$,
CrC$_2$, CrO$_2$, CrMg$_2$, CrP$_2$, CrK$_2$, CrCa$_2$, CrSc$_2$, CrMn$_2$, CrGa$_2$, CrGe$_2$,
CrSe$_2$, CrBr$_2$, CrSr$_2$, CrY$_2$, CrSn$_2$, CrSb$_2$, CrI$_2$, CrCs$_2$, CrPb$_2$, CrBi$_2$,
MnBe$_2$, MnB$_2$, MnO$_2$, MnF$_2$, MnAl$_2$, MnSi$_2$, MnP$_2$, MnS$_2$, MnCl$_2$, MnCa$_2$,
MnSc$_2$, MnCr$_2$, MnMn$_2$, MnSe$_2$, MnBr$_2$, MnIn$_2$, MnSn$_2$, MnOs$_2$, MnAu$_2$, MnPb$_2$,
FeBe$_2$, FeO$_2$, FeF$_2$, FeAl$_2$, FeP$_2$, FeS$_2$, FeGe$_2$, FeAs$_2$, FeSe$_2$, FePd$_2$,
FeSn$_2$, FeBi$_2$, CoO$_2$, CoF$_2$, CoBr$_2$, CoSb$_2$, CoIr$_2$, NiBr$_2$, NiI$_2$, NiIr$_2$,
GaCr$_2$, GaMn$_2$, GaSn$_2$, GeCr$_2$, GeCs$_2$, AsTi$_2$, AsCr$_2$, AsSr$_2$, AsCs$_2$, AsBa$_2$,
RbN$_2$, RbF$_2$, RbCl$_2$, RbBr$_2$, RbI$_2$, SrN$_2$, YF$_2$, YCl$_2$, YBr$_2$, YI$_2$,
ZrCr$_2$, NbF$_2$, MoCl$_2$, MoMn$_2$, MoBr$_2$, MoI$_2$, TcN$_2$, TcF$_2$, TcMn$_2$, TcI$_2$,
RuF$_2$, RuMn$_2$, RhF$_2$, RhI$_2$, PdF$_2$, PdCl$_2$, PdBr$_2$, AgF$_2$, AgCl$_2$, AgMn$_2$,
CdO$_2$, CdCr$_2$, CdY$_2$, InO$_2$, InCr$_2$, InMn$_2$, SnCr$_2$, SnY$_2$, SbK$_2$, SbCa$_2$,
SbSr$_2$, TeSc$_2$, IS$_2$, CsN$_2$, CsF$_2$, CsCr$_2$, CsBr$_2$, CsY$_2$, CsI$_2$, BaN$_2$,
BaCr$_2$, HfY$_2$, TaN$_2$, WBr$_2$, ReB$_2$, ReCs$_2$, OsMn$_2$, OsCs$_2$, OsBa$_2$, IrTi$_2$,
PtF$_2$, PtRu$_2$, AuO$_2$, AuCl$_2$, AuCr$_2$, AuMn$_2$, TlCr$_2$, BiSc$_2$, BiY$_2$.

\section{Comparison with Ref.~\onlinecite{Haastrup_2018}} \label{sec:appendix_comp_database}

Comparison with database in Ref.~\onlinecite{Haastrup_2018} for TMDCs and TMDOs
are summarized in Table~\ref{tab:database_TMDC_TMDO}.

\begin{table}[htb]
\renewcommand{\arraystretch}{1.2}
\newcolumntype{Y}{>{\centering\arraybackslash}p{6.5mm}} 
\centering
       \caption{Comparison with database in Ref.~\onlinecite{Haastrup_2018} for TMDCs and TMDOs.
       Results in Ref.~\onlinecite{Haastrup_2018}  are shown in the left tables.
       Our results are shown in the right tables.
       T, H, T$^\prime$, T$^{\prime\prime}$, M$^\prime$ represent the name of stable structures.
               The left one is more stable than the right one in each part of a cell.}
       \label{tab:database_TMDC_TMDO}
\begin{tabular}{c}
    \begin{minipage}{0.45\linewidth}
       \begin{tabular} {c|*9{Y}} 
           \backslashbox{B}{A}     
   & Sc& Ti& V& Cr& Mn& Fe& Co& Ni& Cu \\
           \hline
   O& TH& T$^\prime$& T$^\prime$& TH& T& T& T& T& T \\
   S& TH& T& HT& H& T$^\prime$& T$^\prime$& T$^\prime$& T& H \\
   Se& TH& T& HT& HT$^\prime$& T& T$^\prime$& T$^\prime$& T& HT \\
   Te& T& T& HT& T& T$^\prime$& T$^\prime$& T$^\prime$& T& T$^\prime$H \\
       \end{tabular}
    \end{minipage}
    \hspace{0.04\linewidth}

    \begin{minipage}{0.45\linewidth}
       \begin{tabular} {c|*9{Y}} 
           \backslashbox{B}{A}     
   & Sc& Ti& V& Cr& Mn& Fe& Co& Ni& Cu \\
           \hline
   O& M$^\prime$& & T$^\prime$& HT& T& T& T& T&  \\
   S& & T& HT& HT& T& TH& T& T&  \\
   Se& & T& HT& T$^\prime$H& T& T$^\prime$H& T& T&  \\
   Te& T& T& T$^\prime$H& & & & & &  \\
       \end{tabular}
    \end{minipage}
\end{tabular}
\vspace{4pt}

\begin{tabular}{c}
    \begin{minipage}{0.45\linewidth}
       \begin{tabular} {c|*9{Y}} 
           \backslashbox{B}{A}     
& Y& Zr& Nb& Mo& Tc& Ru& Rh& Pd& Ag \\
        \hline
O& & T& T$^\prime$H& H& & T$^\prime$& T& T&  \\
S& & T& HT& H& & T$^\prime$& T$^\prime$& T&  \\
Se& & T& HT& H& & T$^\prime$& T$^\prime$& T&  \\
Te& & T& HT& HT$^\prime$& & T$^\prime$& T$^\prime$& T&  \\
       \end{tabular}
    \end{minipage}
    \hspace{0.04\linewidth}

    \begin{minipage}{0.45\linewidth}
       \begin{tabular} {c|*9{Y}} 
           \backslashbox{B}{A}     
& Y& Zr& Nb& Mo& Tc& Ru& Rh& Pd& Ag \\
        \hline
O& & & T$^\prime$H& H& T$^\prime$$^\prime$& T$^\prime$& T& T&  \\
S& & T& HT& H& T$^\prime$$^\prime$& T$^\prime$& T$^\prime$$^\prime$& T&  \\
Se& & T& HT& H& T$^\prime$$^\prime$& T$^\prime$& T$^\prime$$^\prime$& T&  \\
Te& & T& HT$^\prime$& T$^\prime$H& T$^\prime$$^\prime$& T$^\prime$& T$^\prime$$^\prime$& T&  \\
        \end{tabular}
    \end{minipage}
\end{tabular}
\vspace{4pt}

\begin{tabular}{c}
    \begin{minipage}{0.45\linewidth}
       \begin{tabular} {c|*9{Y}} 
           \backslashbox{B}{A}     
& & Hf& Ta& W& Re& Os& Ir& Pt& Au \\
        \hline
O& & T& T$^\prime$H& H& T$^\prime$& T$^\prime$& T$^\prime$& T& T \\
S& & T& HT& H& T$^\prime$& T$^\prime$& T$^\prime$& T& T \\
Se& & T& HT& H& T$^\prime$& T$^\prime$& T$^\prime$& T& T \\
Te& & T& HT& T$^\prime$H& T$^\prime$& T$^\prime$& T$^\prime$& T& T \\
       \end{tabular}
    \end{minipage}
    \hspace{0.04\linewidth}

    \begin{minipage}{0.45\linewidth}
       \begin{tabular} {c|*9{Y}} 
           \backslashbox{B}{A}     
& & Hf& Ta& W& Re& Os& Ir& Pt& Au \\
        \hline
O& & & T$^\prime$H& H& T$^\prime$$^\prime$& & T& T&  \\
S& & T& HT& H& T$^\prime$$^\prime$& T$^\prime$& T$^\prime$$^\prime$& T&  \\
Se& & T& HT& H& T$^\prime$$^\prime$& T$^\prime$& T$^\prime$$^\prime$& T&  \\
Te& & T& M$^\prime$H& T$^\prime$H& T$^\prime$$^\prime$& T$^\prime$& T$^\prime$$^\prime$& T&  \\
        \end{tabular}
    \end{minipage}
\end{tabular}

\renewcommand{\arraystretch}{1.0}
\end{table}

\section{Families of 1T/1H} \label{sec:family_1T1H}

The specific compounds' names of the 1T/1H structures for each combination of groups
in the periodic table are listed in the following subsections.
For example, ``I-XV (2)'' means that atom A and atom B are in group I and XV, respectively. The total number of compounds in the family is written in the parentheses.

\subsection{Families of 1T} \label{sec:family_1T}
\begin{itemize}
   \setlength{\parskip}{0cm}
   \setlength{\itemsep}{0cm}
   \item I-I (2): RbK$_2$, CsCs$_2$
   \item I-XIII (1): LiAl$_2$
   \item I-XV (2): NaN$_2$, KN$_2$
   \item II-I (2): BeRb$_2$, BeCs$_2$
   \item II-II (2): BeCa$_2$, BeBa$_2$
   \item II-TM (1): BeAu$_2$
   \item II-XIV (2): BeGe$_2$, BePb$_2$
   \item II-XVI (1): BeTe$_2$
   \item II-XVII (16): MgF$_2$, MgCl$_2$, MgBr$_2$, MgI$_2$, CaF$_2$, CaCl$_2$, CaBr$_2$, CaI$_2$, SrF$_2$, SrCl$_2$, SrBr$_2$, SrI$_2$, BaF$_2$, BaCl$_2$, BaBr$_2$, BaI$_2$
   \item TM-I (11): CrCs$_2$, PdK$_2$, PdRb$_2$, PdCs$_2$, AgCs$_2$, PtK$_2$, PtRb$_2$, PtCs$_2$, AuK$_2$, AuRb$_2$, AuCs$_2$
   \item TM-II (11): NiBe$_2$, PdCa$_2$, PdSr$_2$, AgCa$_2$, AgSr$_2$, AgBa$_2$, IrSr$_2$, IrBa$_2$, PtSr$_2$, AuCa$_2$, AuSr$_2$
   \item TM-TM (11): MnAu$_2$, NiSc$_2$, NiY$_2$, RhY$_2$, PdSc$_2$, IrSc$_2$, IrY$_2$, IrZr$_2$, PtSc$_2$, PtY$_2$, AuSc$_2$
   \item TM-XII (2): PtHg$_2$, AuHg$_2$
   \item TM-XIII (9): PdTl$_2$, AgAl$_2$, IrIn$_2$, IrTl$_2$, PtGa$_2$, PtIn$_2$, PtTl$_2$, AuAl$_2$, AuIn$_2$
   \item TM-XV (4): TaN$_2$, PtBi$_2$, AuSb$_2$, AuBi$_2$
   \item TM-XVI (39): ScTe$_2$, TiS$_2$, TiSe$_2$, TiTe$_2$, VS$_2$, VSe$_2$, CrO$_2$, CrSe$_2$, MnO$_2$, MnS$_2$, MnSe$_2$, FeO$_2$, FeS$_2$, CoO$_2$, NiO$_2$, NiS$_2$, NiSe$_2$, ZrS$_2$, ZrSe$_2$, ZrTe$_2$, NbS$_2$, NbSe$_2$, RhO$_2$, PdO$_2$, PdS$_2$, PdSe$_2$, PdTe$_2$, HfS$_2$, HfSe$_2$, HfTe$_2$, TaS$_2$, TaSe$_2$, IrO$_2$, PtO$_2$, PtS$_2$, PtSe$_2$, PtTe$_2$, AuSe$_2$, AuTe$_2$
   \item TM-XVII (27): ScBr$_2$, ScI$_2$, VF$_2$, VCl$_2$, VBr$_2$, VI$_2$, MnF$_2$, MnCl$_2$, MnBr$_2$, FeF$_2$, CoF$_2$, CoBr$_2$, NiBr$_2$, NiI$_2$, YF$_2$, YBr$_2$, YI$_2$, TcI$_2$, RuCl$_2$, RuBr$_2$, RhI$_2$, PdF$_2$, PdCl$_2$, PdBr$_2$, OsBr$_2$, OsI$_2$, IrI$_2$
   \item XII-I (1): HgCs$_2$
   \item XII-II (3): CdBa$_2$, HgCa$_2$, HgSr$_2$
   \item XII-XVII (5): CdF$_2$, CdCl$_2$, CdBr$_2$, HgF$_2$, HgCl$_2$
   \item XIII-I (2): BK$_2$, BRb$_2$
   \item XIII-II (2): BMg$_2$, BBa$_2$
   \item XIII-TM (11): BTi$_2$, BNi$_2$, BY$_2$, BNb$_2$, BPd$_2$, BHf$_2$, BTa$_2$, BW$_2$, BIr$_2$, BPt$_2$, InPd$_2$
   \item XIII-XIII (1): AlAl$_2$
   \item XIII-XIV (3): BSi$_2$, BGe$_2$, BSn$_2$
   \item XIII-XVI (2): AlO$_2$, InO$_2$
   \item XIII-XVII (1): TlI$_2$
   \item XIV-I (5): CK$_2$, CRb$_2$, CCs$_2$, SiCs$_2$, GeCs$_2$
   \item XIV-II (17): CBe$_2$, CMg$_2$, CCa$_2$, CBa$_2$, SiCa$_2$, SiSr$_2$, SiBa$_2$, GeMg$_2$, GeCa$_2$, GeSr$_2$, GeBa$_2$, SnCa$_2$, SnSr$_2$, SnBa$_2$, PbCa$_2$, PbSr$_2$, PbBa$_2$
   \item XIV-TM (10): CTi$_2$, CV$_2$, CCr$_2$, CNi$_2$, CY$_2$, CNb$_2$, CPd$_2$, CAg$_2$, CHf$_2$, CTa$_2$
   \item XIV-XII (1): GeCd$_2$
   \item XIV-XIV (3): CGe$_2$, CPb$_2$, SiPb$_2$
   \item XIV-XVI (8): SiO$_2$, SiS$_2$, GeO$_2$, GeSe$_2$, GeTe$_2$, SnO$_2$, SnS$_2$, PbO$_2$
   \item XIV-XVII (9): SiI$_2$, GeCl$_2$, GeBr$_2$, SnBr$_2$, SnI$_2$, PbF$_2$, PbCl$_2$, PbBr$_2$, PbI$_2$
   \item XV-I (10): NLi$_2$, NNa$_2$, NK$_2$, NRb$_2$, NCs$_2$, PK$_2$, PRb$_2$, PCs$_2$, AsCs$_2$, SbK$_2$
   \item XV-II (10): NCa$_2$, NSr$_2$, PCa$_2$, PSr$_2$, PBa$_2$, AsCa$_2$, AsSr$_2$, AsBa$_2$, SbBa$_2$, BiBa$_2$
   \item XV-TM (12): NTi$_2$, NV$_2$, NCr$_2$, NNi$_2$, NY$_2$, NZr$_2$, NNb$_2$, NHf$_2$, NTa$_2$, PFe$_2$, PY$_2$, AsPd$_2$
   \item XV-XIII (2): PTl$_2$, BiAl$_2$
   \item XV-XIV (4): NPb$_2$, PSn$_2$, PPb$_2$, AsSn$_2$
   \item XV-XVI (11): PS$_2$, PSe$_2$, PTe$_2$, AsS$_2$, AsSe$_2$, AsTe$_2$, SbS$_2$, SbSe$_2$, SbTe$_2$, BiS$_2$, BiTe$_2$
   \item XVI-I (20): OLi$_2$, ONa$_2$, OK$_2$, ORb$_2$, OCs$_2$, SLi$_2$, SNa$_2$, SK$_2$, SRb$_2$, SCs$_2$, SeLi$_2$, SeNa$_2$, SeK$_2$, SeRb$_2$, SeCs$_2$, TeLi$_2$, TeNa$_2$, TeK$_2$, TeRb$_2$, TeCs$_2$
   \item XVI-II (8): OMg$_2$, SCa$_2$, SSr$_2$, SBa$_2$, SeCa$_2$, SeSr$_2$, SeBa$_2$, TeSr$_2$
   \item XVI-TM (9): OSc$_2$, OTi$_2$, OV$_2$, OCr$_2$, OFe$_2$, ONb$_2$, SSc$_2$, SY$_2$, SeY$_2$
   \item XVI-XIII (6): OIn$_2$, OTl$_2$, STl$_2$, SeIn$_2$, SeTl$_2$, TeTl$_2$
   \item XVI-XIV (2): SSn$_2$, SPb$_2$
   \item XVI-XVI (5): SeS$_2$, SeSe$_2$, TeS$_2$, TeSe$_2$, TeTe$_2$
   \item XVII-I (10): FNa$_2$, FK$_2$, FRb$_2$, ClK$_2$, ClRb$_2$, ClCs$_2$, BrK$_2$, BrRb$_2$, BrCs$_2$, ICs$_2$
   \item XVII-II (7): ClCa$_2$, ClSr$_2$, ClBa$_2$, BrCa$_2$, BrSr$_2$, BrBa$_2$, ISr$_2$
   \item XVII-TM (1): ClY$_2$
   \item XVII-XII (1): FCd$_2$
   \item XVII-XIII (5): FIn$_2$, FTl$_2$, ClIn$_2$, ClTl$_2$, BrTl$_2$
\end{itemize}

\subsection{Families of 1H} \label{sec:family_1H}
\begin{itemize}
   \setlength{\parskip}{0cm}
   \setlength{\itemsep}{0cm}
   \item I-TM (1): LiAg$_2$
   \item I-XIII (1): LiAl$_2$
   \item I-XV (1): NaN$_2$
   \item I-XVII (5): KF$_2$, KCl$_2$, RbF$_2$, RbCl$_2$, RbBr$_2$
   \item II-II (3): BeBe$_2$, BeCa$_2$, BeBa$_2$
   \item II-TM (4): BeRh$_2$, BePt$_2$, BeAu$_2$, SrAu$_2$
   \item II-XV (4): BeP$_2$, BeSb$_2$, BeBi$_2$, BaN$_2$
   \item II-XVI (7): MgO$_2$, CaO$_2$, CaS$_2$, CaSe$_2$, CaTe$_2$, SrSe$_2$, BaS$_2$
   \item TM-TM (11): FePd$_2$, NiSc$_2$, RuPd$_2$, PdSc$_2$, PdAu$_2$, HfAu$_2$, OsZr$_2$, PtSc$_2$, PtY$_2$, AuSc$_2$, AuY$_2$
   \item TM-XII (1): ScHg$_2$
   \item TM-XIII (2): RhIn$_2$, PdIn$_2$
   \item TM-XV (15): TiN$_2$, VP$_2$, VAs$_2$, MnP$_2$, ZrN$_2$, ZrP$_2$, NbP$_2$, MoSb$_2$, TcN$_2$, TcSb$_2$, TcBi$_2$, HfP$_2$, WAs$_2$, ReN$_2$, OsN$_2$
   \item TM-XVI (24): VS$_2$, VSe$_2$, VTe$_2$, CrO$_2$, CrS$_2$, CrSe$_2$, FeS$_2$, FeSe$_2$, NbO$_2$, NbS$_2$, NbSe$_2$, NbTe$_2$, MoO$_2$, MoS$_2$, MoSe$_2$, MoTe$_2$, TaO$_2$, TaS$_2$, TaSe$_2$, TaTe$_2$, WO$_2$, WS$_2$, WSe$_2$, WTe$_2$
   \item TM-XVII (20): ScF$_2$, ScCl$_2$, ScBr$_2$, ScI$_2$, TiF$_2$, TiCl$_2$, TiBr$_2$, FeF$_2$, YF$_2$, YCl$_2$, YBr$_2$, YI$_2$, ZrF$_2$, ZrCl$_2$, ZrBr$_2$, ZrI$_2$, HfF$_2$, HfCl$_2$, HfBr$_2$, HfI$_2$
   \item XII-II (5): CdCa$_2$, CdBa$_2$, HgCa$_2$, HgSr$_2$, HgBa$_2$
   \item XII-XV (1): CdBi$_2$
   \item XIII-TM (10): BNi$_2$, BRu$_2$, BRh$_2$, BPd$_2$, BW$_2$, BRe$_2$, BIr$_2$, BPt$_2$, AlPd$_2$, AlPt$_2$
   \item XIII-XIII (2): InTl$_2$, TlTl$_2$
   \item XIII-XV (1): AlSb$_2$
   \item XIV-TM (10): CCr$_2$, CMo$_2$, CRu$_2$, CAg$_2$, CW$_2$, CRe$_2$, SiPd$_2$, SiPt$_2$, GePd$_2$, GePt$_2$
   \item XIV-XIV (8): CGe$_2$, CPb$_2$, SiGe$_2$, SiSn$_2$, SiPb$_2$, GeGe$_2$, GeSn$_2$, SnPb$_2$
   \item XV-TM (4): NCr$_2$, NNb$_2$, NTa$_2$, PFe$_2$
   \item XV-XIII (2): PTl$_2$, BiAl$_2$
   \item XV-XIV (3): PGe$_2$, PSn$_2$, PPb$_2$
   \item XVI-II (5): SCa$_2$, SeCa$_2$, SeSr$_2$, TeCa$_2$, TeSr$_2$
   \item XVI-TM (9): OTi$_2$, OV$_2$, OCr$_2$, OFe$_2$, OZr$_2$, ONb$_2$, SSc$_2$, SeSc$_2$, SeY$_2$
   \item XVI-XIII (6): OIn$_2$, OTl$_2$, STl$_2$, SeIn$_2$, SeTl$_2$, TeTl$_2$
   \item XVI-XIV (1): SSn$_2$
   \item XVII-I (7): FNa$_2$, FK$_2$, ClK$_2$, ClRb$_2$, BrK$_2$, BrCs$_2$, IK$_2$
   \item XVII-II (7): ClCa$_2$, ClSr$_2$, ClBa$_2$, BrCa$_2$, BrSr$_2$, BrBa$_2$, ISr$_2$
   \item XVII-TM (1): ClY$_2$
   \item XVII-XII (1): FCd$_2$
   \item XVII-XIII (3): FIn$_2$, FTl$_2$, ClIn$_2$
\end{itemize}

\section{NEB results for memory structures} \label{sec:appendix_memory_neb}

\begin{figure}[htb]
\includegraphics[width=0.98\linewidth]{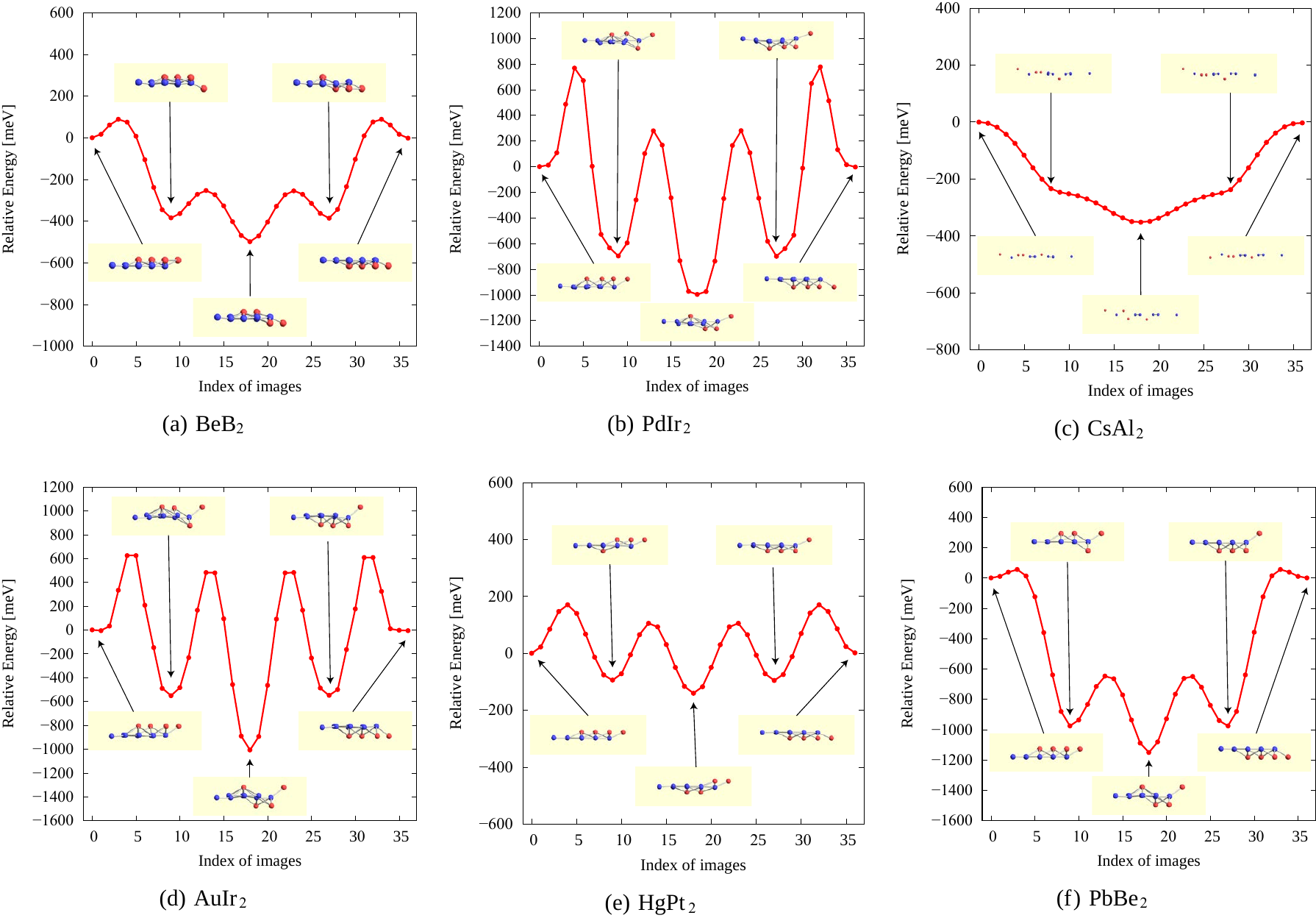}
\caption{The NEB results for memory structures:
        (a) BeB$_2$, (b) PdIr$_2$, (c) CsAl$_2$, (d) AuIr$_2$, (e) HgPt$_2$, and (f) PbBe$_2$.
}\label{fig:NEB_memory}
\end{figure}

NEB results for memory structures such as BeB, PdIr, CsAl, AuIr, HgPt, and PbBe
are shown in Fig.~\ref{fig:NEB_memory}.
For each compounds, the five structures on the local minima
are prepared to represent a model of a binary digit storage application.
Each minimum energy path between two of them is obtained by optimizing 8 images on the path.
The whole energy path is obtained by connecting all the minimum energy paths.
The corresponding geometrical structures of images on the local energy minima
are depicted in Fig.~\ref{fig:NEB_memory}.

\end{document}